\documentclass[submission, Phys]{SciPost}

\usepackage[utf8]{inputenc} 
\usepackage[T1]{fontenc} 	
\usepackage[english]{babel} 

\usepackage[bitstream-charter]{mathdesign}
\usepackage{geometry} 		
\usepackage{mathtools} 		
\usepackage{float} 			
\usepackage{graphicx} 		
\usepackage{tabularx} 		
\usepackage{booktabs} 		
\usepackage{color, xcolor} 	
\usepackage{pdfpages} 		
\usepackage{extarrows} 		
\usepackage{multirow} 		
\usepackage{multicol} 		
\usepackage{enumitem} 		
\usepackage{xspace} 		
\usepackage{stackrel} 		
\usepackage{tikz} 			
\usepackage{braket} 		
\usepackage{bm} 			
\usepackage{tensor} 		
\usepackage{slashed} 		
\usepackage{siunitx} 		
\usepackage{lastpage} 		
\usepackage{cite} 			
\usepackage[normalem]{ulem} 
\usepackage{fontawesome} 	
\usepackage{tocloft} 		
\usepackage{titlesec} 		
\usepackage{doi} 			
\usepackage{hyperref} 		
\usepackage[most]{tcolorbox} 					
\usepackage[nameinlink, capitalize]{cleveref} 	
\usepackage[nottoc, notlot, notlof]{tocbibind} 	
\usepackage[ruled, vlined]{algorithm2e} 		
\usepackage{makecell}
\usepackage[makeroom]{cancel}
\usepackage{feynmf}

\makeatletter
\def\BState{\State\hskip-\ALG@thistlm}
\makeatother

\makeatletter
\@ifundefined{pdfoutput}{}{\DeclareGraphicsRule{*}{mps}{*}{}}
\makeatother

\makeatletter
\DeclareRobustCommand*{\bfseries}{%
   \not@math@alphabet\bfseries\mathbf
   \fontseries\bfdefault\selectfont
   \boldmath
}
\makeatother

\hypersetup{
	pdftitle={NCVs for LO and NLO},
	pdfauthor={Heimel et al.},
	colorlinks=true, 			
	linkcolor={red!50!black}, 	
	citecolor={blue!50!black}, 	
	urlcolor={blue!80!black} 	
} 

\DeclareSymbolFont{usualmathcal}{OMS}{cmsy}{m}{n}
\DeclareSymbolFontAlphabet{\mathcal}{usualmathcal}

\SetArgSty{textnormal}
\SetKwComment{Comment}{{\small\#}~}{}
\SetCommentSty{mycommfont}

\setitemize{itemsep=0pt, parsep=0pt} 				
\setenumerate{itemsep=0pt, parsep=0pt} 				
\setitemize{itemsep=2pt,topsep=2pt,parsep=0pt,partopsep=0pt,leftmargin=*}
\setenumerate{itemsep=0pt,topsep=2pt,parsep=0pt,partopsep=0pt,labelindent=3pt,leftmargin=*}
\newlist{todolist}{itemize}{2}
\setlist[todolist]{label=$\square$}
\usepackage{pifont}

\usepackage{amsmath}
 
\usepackage{amsthm} 		
\theoremstyle{definition}

\definecolor{red_cb}{HTML}{e41a1c}
\definecolor{blue_cb}{HTML}{377eb8}
\definecolor{green_cb}{HTML}{4daf4a}
\definecolor{purple_cb}{HTML}{984ea3}
\definecolor{orange_cb}{HTML}{ff7f00}

\definecolor{EmeraldGreen}{HTML}{1ea78d}
\definecolor{EnglishRed}{HTML}{b02427}
\hypersetup{colorlinks=true,urlcolor=EmeraldGreen,citecolor=EmeraldGreen,linkcolor=EnglishRed}

\newcommand{\eg}{\text{e.g.}\;}
\newcommand{\ie}{\text{i.e.}\;}

\newcommand{\eqperiod}{\quad\text{.}} 	

\newcommand{\mwith}{\text{with}}
\newcommand{\mand}{\text{and}}

\newcommand{\jcuts}{J}

\newcommand{\XLangle}{\Bigl\langle}
\newcommand{\XRangle}{\Bigr\rangle}

\newcommand{\XXXLangle}{\Biggl\langle}
\newcommand{\XXXRangle}{\Biggr\rangle}

\newcommand{\qqquad}{\qquad\quad}
\newcommand{\qqqquad}{\qquad\qquad}

\def\dd{\mathrm{d}}

\newcommand\one{\leavevmode\hbox{\small1\normalsize\kern-.33em1}}
\newcommand{\euler}{\mathrm{e}} 			

\newcommand{\loss}{\mathcal{L}} 	

\newcommand{\vegas}{\texttt{Vegas}\xspace}

\newcommand{\mg}{\texttt{MG5aMC}\xspace}

\newcommand{\sherpa}{\texttt{Sherpa}\xspace}
\newcommand{\madnis}{\texttt{MadNIS}\xspace}

\newcommand{\madspace}{\texttt{MadSpace}\xspace}
\newcommand{\cudacpp}{\texttt{CUDACPP}\xspace}

\newcommand{\arXiv}[2][]{%
	\ifthenelse{\equal{#1}{}}%
	{\href{http://arxiv.org/abs/#2}{arXiv:#2}}%
	{\href{http://arxiv.org/abs/#2}{arXiv:#2~[#1]}}}

\def\slashchar#1{\setbox0=\hbox{$#1$}           
   \dimen0=\wd0                                 
   \setbox1=\hbox{/} \dimen1=\wd1               
   \ifdim\dimen0>\dimen1                        
      \rlap{\hbox to \dimen0{\hfil/\hfil}}      
      #1                                        
   \else                                        
      \rlap{\hbox to \dimen1{\hfil$#1$\hfil}}   
      /                                         
   \fi}

\newcommand{\tikznode}[2]{%
\ifmmode%
\tikz[remember picture,baseline=(#1.base),inner sep=0pt] \node (#1) {$#2$};%
\else
\tikz[remember picture,baseline=(#1.base),inner sep=0pt] \node (#1) {#2};%
\fi}

\def\mathswitchr#1{\relax\ifmmode{\mathrm{#1}}\else$\mathrm{#1}$\xspace\fi}
\def\mathswitch#1{\relax\ifmmode#1\else$#1$\xspace\fi}

\newcommand{\Pq}{\mathswitch q}
\newcommand{\Pqbar}{\mathswitch{\bar q}}

\newcommand{\Pg}{\mathswitchr g}

\newcommand{\Pep}{\mathswitchr {e^+}}
\newcommand{\Pem}{\mathswitchr {e^-}}

\newcommand{\Pd}{\mathswitchr d}
\newcommand{\Pu}{\mathswitchr u}
\newcommand{\Ps}{\mathswitchr s}
\newcommand{\Pc}{\mathswitchr c}

\newcommand{\Pt}{\mathswitchr t}

\newcommand{\Ptbar}{\mathswitchr{\bar t}}

\newcommand{\Mt}{\mathswitch {m_\Pt}}

\graphicspath{{./figs/}}

\begin{document}

\begin{center}{\Large \textbf{\color{scipostdeepblue}{
Neural Control Variates at LO and NLO
}}}
\end{center}

\begin{center}
Theo Heimel\textsuperscript{1},
Tilman Plehn\textsuperscript{2,3},
Rebecca Revelli\textsuperscript{2}, 
Sophia Vent\textsuperscript{2},
and Ramon Winterhalder\textsuperscript{4}
\end{center}

\begin{center}
{\bf 1} CP3, Universit\'e catholique de Louvain, Louvain-la-Neuve, Belgium
\\
{\bf 2} Institut f\"ur Theoretische Physik, Universit\"at Heidelberg, Germany
\\
{\bf 3} Interdisciplinary Center for Scientific Computing (IWR), Universit\"at Heidelberg, Germany \\
{\bf 4} TIFLab, Universit\`a degli Studi di Milano \& INFN Sezione di Milano, Italy
\end{center}

\begin{center}
\today
\end{center}

\section*{\color{scipostdeepblue}{Abstract}}
{\bf We employ neural control variates to minimize the range of event weights and avoid negative weights for phase-space integration and event generation. A signed control variate, built from two normalizing flows, fulfills both tasks. Combined with neural importance sampling, it significantly reduces the computational cost of LO and NLO predictions. For the NLO case, our conditional neural control variate can be viewed as a trainable subtraction term, complementing the established physics subtraction schemes  for enhanced sampling performance.}

\vspace{1pt}
\noindent\rule{\textwidth}{1pt}
\tableofcontents\thispagestyle{fancy}
\noindent\rule{\textwidth}{1pt}
\vspace{1pt}

\clearpage
\section{Introduction}
\label{sec:intro}

Precise and fast first-principle simulations provide the crucial theory ingredient to the LHC program~\cite{Campbell:2022qmc}. Event generators~\cite{Bierlich:2022pfr,Sherpa:2024mfk,Bellm:2025pcw,Maltoni:2002qb,Alwall:2007st,Alwall:2011uj,Alwall:2014hca,Frederix:2018nkq} provide the interpretation of every LHC analysis. They combine perturbative QCD calculations with parton showers and hadronization and allow us to understand the data in terms of theory. For the upcoming high-luminosity LHC, higher-order predictions in the QCD and electroweak couplings are indispensable. 

Improving event generators and eliminating critical bottlenecks is the key task of modern machine learning (ML) in theoretical particle physics~\cite{Butter:2022rso,Plehn:2022ftl,Ubiali:2026myh}. Neural networks have been shown to speed up amplitude calculations~\cite{Bishara:2019iwh,Badger:2020uow,Aylett-Bullock:2021hmo,Maitre:2021uaa,Danziger:2021eeg,Winterhalder:2021ngy,Janssen:2023ahv,Maitre:2023dqz,Brehmer:2024yqw,Breso-Pla:2024pda,Herrmann:2025nnz,Favaro:2025pgz,Villadamigo:2025our,Bahl:2026jvt} including calibrated learned uncertainties~\cite{Badger:2022hwf,Bahl:2024gyt,Bahl:2025xvx,Beccatini:2025tpk,Bahl:2026qaf}, improve hadronization~\cite{Ilten:2022jfm,Ghosh:2022zdz,Chan:2023ume,Bierlich:2023zzd,Chan:2023icm,Bierlich:2024xzg,Assi:2025avy,Butter:2025wxn}, and generate complete collider events~\cite{Hashemi:2019fkn,DiSipio:2019imz,Butter:2019cae,Alanazi:2020klf,Butter:2021csz,Butter:2023fov,Quetant:2024ftg,Brehmer:2024yqw,Butter:2024zbd,Favaro:2025pgz,Bahl:2025ryd,Bahl:2026lsa}, while agentic systems increasingly support the standard simulation and analysis tools~\cite{Diefenbacher:2025zzn,Bakshi:2025fgx,Gendreau-Distler:2025fsj,Menzo:2025cim,Plehn:2026gxv,Esmail:2026jpb,Qiu:2026iby,Birk:2026zpd,Desai:2026nmx,Costa:2026oew,Diefenbacher:2026azr}.
Neural importance sampling (NIS)~\cite{Bendavid:2017zhk,Klimek:2018mza,Chen:2020nfb,Gao:2020vdv,Deutschmann:2024lml} has been successfully applied using \madnis~\cite{Heimel:2022wyj,Heimel:2023ngj,Heimel:2024wph} and its \sherpa counterpart~\cite{Gao:2020zvv, Bothmann:2020ywa,Bothmann:2023siu,Bothmann:2025lwg,Bothmann:2026dar}. It targets the wide range of event weights, especially from kinematic tails, which leads to a large variance and poor unweighting efficiency.

Recently, normalizing-flow samplers have been extended to NLO~\cite{Gao:2020zvv} and NNLO~\cite{Janssen:2025zke} and have been combined with amplitude surrogates for virtual and real corrections to a NLO-\madnis framework~\cite{DeCrescenzo:2026tsp}. These studies raise the question whether NIS methods are numerically efficient in the presence of subtraction schemes that regularize real emission~\cite{Catani:1996vz,Catani:2002hc,Frixione:1995ms,Frederix:2009yq}. Going beyond a pure regularization, subtraction terms are often used to shift non-trivial phase-space integration from the real-emission phase space to the Born-like phase space. Here, they aim for small and featureless subtracted integrands, at the risk of creating negative event weights with a catastrophic effect on variance reduction and unweighting.

To $(i)$ address large event weight ranges and $(ii)$ reduce negative event weights, we propose to use neural control variates (NCVs)~\cite{neuralcontrol}.  They replace the ratio of integrand and sampling density typical for importance sampling with a numerically more stable subtraction. Control variates and importance sampling are complementary and can be combined as classical algorithms~\cite{Shyamsundar:2023jtz}. Their combination can be expanded beyond just integration. The NCV contribution can be sampled trivially, combining a normalizing flow with a learned integral value.  Extending NCVs to real emission corrections can be viewed as learnable subtraction, reducing the physics-defined subtraction to their original purpose of removing soft and collinear divergences.

We introduce the NCV in Section~\ref{sec:method} with two objectives, reducing the range of event weights and removing negative event weights. We illustrate its conditional and unconditional variants in a simple toy example. In Section~\ref{sec:lo}, we first target the weight range at LO with a combined NCV-NIS setup and benchmark it for top pair production with two gluons. In Section~\ref{sec:nlo}, we then target negative weights at NLO. A conditional NCV stabilizes the real-emission integration, while an unconditional NCV controls the Born-like integral. We apply this combined setup to top pair production with a hard jet and to three-jet production in electron-positron scattering, showing the same significant improvements.

\clearpage
\section{Neural control variates}
\label{sec:method}

We introduce neural control variates (NCVs) as a generic tool for Monte Carlo integration and sampling. A single signed control variate is subtracted from the integrand. It is built from two components that target the positive and negative parts of the integrand. Together, they serve two purposes: reducing the range of event weights and removing negative weights.

\subsection{Integration and sampling}
\label{sec:method_basic}

A control variate is a function with a known integral that we subtract from an integrand to reduce the variance of a Monte Carlo estimate. We work on the unit hypercube $x\in[0,1]^d$, as produced by the phase-space mappings. We want to integrate over a function $f(x)$ with variable sign,
\begin{align}
    \sigma
    = \int \dd x \,f(x) \; .
\label{eq:cv_generic_integral}
\end{align}
We can then subtract and compensate any function $c(x)$ with a known integral~\cite{Shyamsundar:2023jtz},
\begin{align}
    \sigma
    &= \int \dd x \, \Big[ f(x) - c(x) \Big]
    + \int \dd x \, c(x) \; .
    \label{eq:cv_generic}
\end{align}
We construct a neural control variate $c_\theta$, with trainable weights $\theta$, from two positive components, each the product of a normalizing flow and a learned scalar normalization,
\begin{align}
    c_\theta(x)
    = C_{+,\theta}\,p_{+,\theta}(x) - C_{-,\theta}\,p_{-,\theta}(x)
    \qquad \mwith \qquad
    C_{\pm,\theta}\ge0\;,
    \qquad
    \int \dd x\; p_{\pm,\theta}(x) = 1 \; .
    \label{eq:cv_two_components}
\end{align}
The integrated control variate is the difference of the two learned normalizations,
\begin{align}
    \int \dd x\; c_\theta(x) &= C_{+,\theta}-C_{-,\theta} \equiv C_\theta \notag \\
    \Rightarrow \quad
    \sigma
    &= \int \dd x \,\Big[ f(x) - c_\theta(x) \Big] + C_\theta
    \equiv \sigma_{\text{res},\theta} + C_\theta \; .
    \label{eq:ps_cv0}
\end{align}
The key observation for event generation is that the compensating term can also be sampled. A learned importance-sampling density $g_\phi(x)$, with trainable weights $\phi$, gives us
\begin{align}
\sigma
&=
\left\langle \frac{f(x) - c_\theta(x)}{g_\phi(x)} \right\rangle_{x\sim g_\phi}
+ C_{+,\theta}\,\XLangle 1 \XRangle_{x\sim p_{+,\theta}}
- C_{-,\theta}\,\XLangle 1 \XRangle_{x\sim p_{-,\theta}}
\notag \\
&\equiv
\big\langle w_{\text{res},\phi,\theta}(x)\big\rangle_{x\sim g_\phi}
+ C_\theta
\qqqquad \mwith \qquad
w_{\text{res},\phi,\theta}(x) = \frac{f(x) - c_\theta(x)}{g_\phi(x)}\;.
\label{eq:cv_mc}
\end{align}
The two compensating terms are sampled directly, their events carry unit weight $\pm1$ and are unweighted with perfect efficiency. The residual $\sigma_{\text{res},\theta}$ requires evaluations of $f(x)$ and an unweighting step. The goal of efficient event generation is to absorb as much of the integral as possible into the trivial term, while keeping the residual integrand small and positive.

Each NCV component has a positive target, the corresponding sign part of the integrand,
\begin{align}
    f^+(x) = \max\big(0,f(x)\big)
    \qquad \mand \qquad
    f^-(x) = \max\big(0,-f(x)\big)\; ,
\label{eq:target_pm}
\end{align}
so the optimum of each component is the contribution and the shape of its target,
\begin{align}
C_{\pm,\theta}^{\rm opt}=\int\dd x\,f^\pm(x)
\qquad \mand \qquad
p_{\pm,\theta}^{\rm opt}(x) = \frac{f^\pm(x)}{C_{\pm,\theta}^{\rm opt}}\;.
\label{eq:target_optimum}
\end{align}
In that case, we obtain $c_\theta = f^+ - f^- = f$, the residual vanishes everywhere, and the entire cross section is given by the trivially sampled $C_\theta$. This makes our goals explicit:
\begin{enumerate}
\item \textbf{Range reduction:} the positive component subtracts the bulk of $f>0$, compressing the range of the residual event weights. It must not overshoot, as $C_{+,\theta}\,p_{+,\theta}>f$ would drive the residual negative.
\item \textbf{Sign removal:} the negative component adds $C_{-,\theta}\,p_{-,\theta}$ where $f<0$, lifting the residual to its positive part. Here, a slight overshoot is benign, and in fact required for strict positivity.
\end{enumerate}
For a positive integrand, $f^- = 0$ implies $C_{-,\theta}=0$ and only the positive component is active. This happens for LO event generation in Section~\ref{sec:lo}, where the NCV is a single subtracted flow.

\subsubsection*{Nested control variates}
\label{sec:method_nested}

Equation~\eqref{eq:ps_cv0} defines an NCV acting on the integrand $f(x)$, irrespective of its dimension or internal structure. However, we are often interested in cases where the variables factorize into two groups, $x=(x_B,x_R)$, and only part of the integrand depends on both,
\begin{align}
f(x_B,x_R) = b(x_B) + r(x_B,x_R)\;.
\label{eq:nested_split}
\end{align}
A single NCV on the joint space does not exploit this structure. Instead, we proceed in two steps, mirroring the structure of the integrand. First, a conditional NCV acts on $r$ for fixed $x_B$. It is built exactly as in Eq.\eqref{eq:cv_two_components}, but from conditional flows with $x_B$-dependent normalizations,
\begin{align}
c_\eta(x_R|x_B)
&= A_{+,\eta}(x_B)\,p_{+,\eta}(x_R|x_B) - A_{-,\eta}(x_B)\,p_{-,\eta}(x_R|x_B)\;,
\notag \\
\mwith \qquad
&A_{\pm,\eta}(x_B)\ge0
\qquad \mand \qquad
\int \dd x_R\;p_{\pm,\eta}(x_R|x_B)=1\;.
\label{eq:cv_conditional_def}
\end{align}
Its integral over $x_R$ is known,
\begin{align}
\int \dd x_R \; c_\eta(x_R|x_B)
= A_{+,\eta}(x_B)-A_{-,\eta}(x_B) \equiv A_\eta(x_B)\;,
\label{eq:cv_conditional_mass}
\end{align}
so subtracting it and compensating gives
\begin{align}
\sigma
&=
\int \dd x_B\;
\Big[ \underbrace{ b(x_B) + A_\eta(x_B) }_{\textstyle \equiv b_{\text{res},\eta}(x_B)} \Big]
+
\int \dd x_B\,\dd x_R\;
\Big[ \underbrace{r(x_B,x_R) - c_\eta(x_R|x_B) }_{\textstyle \equiv r_{\text{res},\eta}(x_B,x_R)} \Big]\; .
\label{eq:nested_cv_cond}
\end{align}
The conditional NCV flattens the integrand in $x_R$ and removes its sign changes, while $A_\eta$ is shifted into the $x_B$ integration. 
A second, unconditional NCV then acts as in Eq.\eqref{eq:cv_two_components},
\begin{align}
\sigma
&=
\int \dd x_B\;
\Big[
\underbrace{b_{\text{res},\eta}(x_B) - c_\theta(x_B)}_{\textstyle \equiv b_{\text{res},\theta,\eta}(x_B)}
\Big]
+ C_\theta
+
\int \dd x_B\,\dd x_R\;
r_{\text{res},\eta}(x_B,x_R) \; .
\label{eq:nested_cv_uncond}
\end{align}
The targets follow the same pattern as in Eq.\eqref{eq:target_pm},
\begin{alignat}{4}
\text{conditional}
&\qqqquad
&& r^\pm(x_B,x_R)
&& \qquad\Rightarrow\qquad
& A^{\rm opt}_{\pm,\eta}(x_B)
&= \int \dd x_R\; r^\pm(x_B,x_R)\;,
\notag\\
\text{unconditional}
&\qqqquad
&& b^\pm_{\text{res},\eta}(x_B)
&& \qquad\Rightarrow\qquad
& C^{\rm opt}_{\pm,\theta}
&= \int \dd x_B\; b^\pm_{\text{res},\eta}(x_B)\;.
\label{eq:nested_targets}
\end{alignat}
The second target is defined using the subtracted integrand, so the two NCVs do not compete. Ideally, both residuals vanish and the whole integral is given by the $C_\theta$ term. If $r_{\text{res},\eta}\ge0$ and $b_{\text{res},\theta,\eta}\ge0$, the residual sample involves only positive weights and $C_\theta$ does not generate an variance. For sampling, the two residuals are combined into a single residual integrand,
\begin{align}
f_{\text{res},\theta,\eta}(x_B,x_R)
= b_{\text{res},\theta,\eta}(x_B)
+  r_{\text{res},\eta}(x_B,x_R)\; .
\label{eq:nested_residual}
\end{align}
It is sampled by a single NIS flow $g_\phi(x_B,x_R)$.

\subsection{Training the NCV}
\label{sec:method_loss}

Directly minimizing the variance of the residual weights $w_{\text{res},\phi,\theta}$ of Eq.\eqref{eq:cv_mc} supplemented by a positivity constraint would in principle converge to Eq.\eqref{eq:target_optimum}. However, for integrands with deep and narrow structures this does not converge well, so we learn the shape of each NCV flow and its normalization separately, supplemented by the NIS loss and by a penalty for negative weights~\cite{neuralcontrol},
\begin{align}
\loss =
\sum_{s=\pm}\Big[\loss_{\text{shape},s} + \lambda_{\text{int}}\,\loss_{\text{int},s}\Big]
+ \loss_{\text{NIS},\phi}
+ \lambda_{\text{neg}}\,\loss_{\text{neg},\theta}\; .
\label{eq:loss_single_ncv_total}
\end{align}
We train all networks simultaneously and evaluate the losses on joint samples,
\begin{align}
q(x)
=
\frac{g_\phi(x) + p_{+,\theta}(x) + p_{-,\theta}(x)}{3}\Bigg|_{\text{no-grad}}\; .
\label{eq:method_mixture}
\end{align}
This gives us the two target weights and the residual weight
\begin{align}
w_{\pm}(x) = \frac{f^\pm(x)}{q(x)}
\qquad \mand \qquad
w_\text{NIS}(x) = \frac{f(x)-c_\theta(x)\big|_{\text{no-grad}}}{q(x)}\;.
\label{eq:def_weights}
\end{align}
We also need normalizations that make each term invariant under a rescaling $f\to \kappa f$, so that $\lambda_\text{int}$ and $\lambda_\text{neg}$ carry over between applications with very different cross sections. For the shape and integral losses this is fixed by
\begin{align}
Z_\pm = C_{\pm,\theta}\Big|_{\text{no-grad}}\;.
\label{eq:normalization}
\end{align}
For the sampler and the negativity barrier, we denote the normalizations by $Z_\text{NIS}$ and $Z_\text{neg}$ and treat them as conventions rather than derived quantities. $Z_\text{NIS}$ is a global factor of a weighted log-likelihood, does not affect its minimization, and only normalizes this loss term relative to others.  $Z_\text{neg}$ enters inside a softplus function, and only the ratio $\beta_\text{max}/Z_\text{neg}$ is meaningful. A natural candidate is $C_{+,\theta}+C_{-,\theta}$, which remains positive even when $C_\theta$ passes through zero for strong cancellations. Alternatively, we can choose a single component or the absolute integral of the residual itself. The individual loss terms in Eq.\eqref{eq:loss_single_ncv_total} then read
\begin{enumerate}
    \item shape loss for each NCV flow $p_{\pm,\theta}$ as a weighted log-likelihood,
    \begin{align}
    \loss_{\text{shape},\pm}
    &=
    -\frac{1}{Z_\pm}
    \left\langle
        w_\pm(x)\,
        \log p_{\pm,\theta}(x)
    \right\rangle_{x\sim q}\;.
    \label{eq:loss_shape_single_ncv}
    \end{align}

    \item integral loss for each learned normalization $C_{\pm,\theta}$,
    \begin{align}
    \loss_{\text{int},\pm}
    &=
    \frac{1}{Z^2_\pm}
    \left[
        C_{\pm,\theta} -
        \big\langle w_\pm(x) \big\rangle_{x\sim q}
    \right]^2\;.
    \label{eq:loss_norm_single_ncv}
    \end{align}

    \item weighted NIS log-likelihood loss for $g_\phi$, trained on the residual integrand,
    \begin{align}
    \loss_{\text{NIS},\phi}
    &=
    -\frac{1}{Z_\text{NIS}}
    \left\langle
    |w_\text{NIS}(x)|\,\log g_\phi(x)
    \right\rangle_{x\sim q}\;.
    \label{eq:loss_is_single_ncv}
    \end{align}

    \item softplus barrier regularizing the residual, as the NCV flows only approximate their targets,
    \begin{align}
    \loss_{\text{neg},\theta}
    &=
    \left\langle \Psi\!\left(\frac{f(x)- c_\theta(x)}{Z_\text{neg}\,q(x)}\right)\right\rangle_{x\sim q}
    \notag\\
    \mwith &\qquad
    \Psi(t)=\log\!\big(1+\euler^{-\beta(t)\cdot t}\big)
    \qquad \mand \qquad
    \beta(t) =  \min\left\{\beta_{\text{max}},\; \frac{100\, t}{N_\text{epochs}}\right\}\;,
    \label{eq:positivity_barriere}
    \end{align}
    with a linearly annealed barrier scale $\beta(t)$.
    Initially, the NCVs focus on reproducing their targets, and then the barrier removes the remaining negative weights.
\end{enumerate}
%

\subsubsection*{Nested training}
\label{sec:training_nested}

The nested construction contains two NCVs. They are trained with the same building blocks as the single NCV. The two NCVs live in different spaces, so we use two mixtures. The conditional NCV and the sampler act on the full space and are evaluated on
\begin{align}
q(x_B,x_R)
&=
\frac{1}{4}\Big[\,
g_\phi(x_B,x_R)
+ \sum_{s=\pm} p_{s,\theta}(x_B)\,p_{s,\eta}(x_R|x_B)
+ p_U(x_B,x_R)
\,\Big]\Bigg|_{\text{no-grad}} \; .
\label{eq:nested_mixture_joint}
\end{align}
The unconditional NCV is a function of $x_B$ alone and is evaluated on a
separate mixture,
\begin{align}
q(x_B)
&=
\frac{1}{3}\Big[\,
p_U(x_B) + p_{+,\theta}(x_B) + p_{-,\theta}(x_B)
\,\Big]\Bigg|_{\text{no-grad}} \; .
\label{eq:nested_mixture_born}
\end{align}
Both mixtures are known exactly, and every estimator below is unbiased.  The three weights entering the losses are
\begin{align}
w_{\pm,r}(x_B,x_R) &= \frac{r^\pm(x_B,x_R)}{q(x_B,x_R)} \notag \\
w_{\pm,b}(x_B) &= \frac{b^\pm_{\text{res},\eta}(x_B)\big|_{\text{no-grad}}}{q(x_B)} \notag\\
w_\text{NIS}(x_B,x_R) &= \frac{f_{\text{res},\theta,\eta}(x_B,x_R)\big|_{\text{no-grad}}}{q(x_B,x_R)}\;,
\label{eq:nested_weights}
\end{align}
with normalizations
\begin{align}
Z_{\pm,A} = \left\langle \frac{A_{\pm,\eta}(x_B)\big|_{\text{no-grad}}}{q(x_B)}\right\rangle_{x_B\sim q(x_B)}
\qquad \mand \qquad
Z_{\pm,C} = C_{\pm,\theta}\Big|_{\text{no-grad}}\; .
\label{eq:nested_normalizations}
\end{align}
The normalizations $Z_\text{NIS}$, $Z_{\text{neg},r}$ and $Z_{\text{neg},b}$ follow the same convention as in the single-NCV case. The individual loss terms read
\begin{enumerate}
\item shape losses for the two conditional flows as weighted log-likelihoods at fixed $x_B$,
\begin{align}
\loss_{\text{shape},\pm,\eta}
&=
-\frac{1}{Z_{\pm,A}}
\Big\langle
w_{\pm,r}(x_B,x_R)\,\log p_{\pm,\eta}(x_R|x_B)
\Big\rangle_{x\sim q(x_B,x_R)}\; .
\label{eq:loss_shape_cond}
\end{align}

\item integral losses for the two conditional normalizations, with a regression of
$A_{\pm,\eta}(x_B)$ onto the integral of its target
\begin{align}
\loss_{\text{int},\pm,\eta}
&=
\frac{1}{Z^2_{\pm,A}}
\left\langle
\frac{\big[A_{\pm,\eta}(x_B)-r^\pm(x_B,x_R)\big]^2}{q(x_B,x_R)}
\right\rangle_{x\sim q(x_B,x_R)}\; .
\label{eq:loss_int_cond}
\end{align}

\item shape and integral losses for the two unconditional flows and normalizations, as in Eqs.\eqref{eq:loss_shape_single_ncv} and~\eqref{eq:loss_norm_single_ncv},
\begin{align}
\loss_{\text{shape},\pm,\theta}
&=
-\frac{1}{Z_{\pm,C}}
\Big\langle
w_{\pm,b}(x_B)\,\log p_{\pm,\theta}(x_B)
\Big\rangle_{x_B\sim q(x_B)}\;,
\notag\\
\loss_{\text{int},\pm,\theta}
&=
\frac{1}{Z^2_{\pm,C}}
\Big[\,C_{\pm,\theta}-\big\langle w_{\pm,b}(x_B)\big\rangle_{x_B\sim q(x_B)}\,\Big]^2\;;
\label{eq:loss_uncond}
\end{align}

\item NIS log-likelihood loss, evaluated on the combined residual of Eq.\eqref{eq:nested_residual},
\begin{align}
\loss_{\text{NIS},\phi}
&=
-\frac{1}{Z_\text{NIS}}
\Big\langle
\big|w_\text{NIS}(x_B,x_R)\big|\,\log g_\phi(x_B,x_R)
\Big\rangle_{x\sim q(x_B,x_R)}\; .
\label{eq:loss_nis_nested}
\end{align}

\item two softplus barriers, one for each residual,
\begin{align}
\loss_{\text{neg}}
&=
\left\langle
\Psi\!\left(\frac{r_{\text{res},\eta}(x_B,x_R)}{Z_{\text{neg},r}\,q(x_B,x_R)}\right)
\right\rangle_{x\sim q(x_B,x_R)}
+
\left\langle
\Psi\!\left(\frac{b_{\text{res},\theta,\eta}(x_B)}{Z_{\text{neg},b}\,q(x_B)}\right)
\right\rangle_{x_B\sim q(x_B)}\;.
\label{eq:loss_neg_nested}
\end{align}
Unlike in Eq.\eqref{eq:nested_weights}, where $A_{\pm,\eta}$ is detached so that the two control variates do not compete, this barrier retains the full $A_{\pm,\eta}$ dependence of $b_{\text{res},\theta,\eta}=b+A_\eta-c_\theta$. This couples the two terms. The normalization $A_{\pm,\eta}$ enters the two residuals with opposite signs,
\begin{align}
\frac{\partial\, r_{\text{res},\eta}}{\partial A_{\pm,\eta}}
= \mp\, p_{\pm,\eta}(x_R|x_B)
\qquad \mand \qquad
\frac{\partial\, b_{\text{res},\theta,\eta}}{\partial A_{\pm,\eta}}
= \pm 1 \; .
\end{align}
The derivative of the sum of the two barriers with respect to $A_{\pm,\eta}$ is that of $f_{\text{res},\theta,\eta}$, while $C_{\pm,\theta}$ appears in the second term only. Positivity is imposed on the combined residual along the direction the two pieces share, which is the correct condition because the combined residual determines the sign of the sampled event weights. Demanding each piece to be positive is not necessary and drives $|A_\eta|$ and $C_\theta$ to large values. At the same time, the two terms are not equivalent to a single barrier on $f_{\text{res},\theta,\eta}$, since $\Psi(a)+\Psi(b)\neq\Psi(a+b)$.
\end{enumerate}
The complete nested loss is the sum of these terms,
\begin{align}
\loss_\text{nested}
&=
\underbrace{\sum_{s=\pm}\Big[\loss_{\text{shape},s,\eta} + \lambda_{\text{int}}\,\loss_{\text{int},s,\eta}\Big]}_{\text{conditional NCV}}
\notag\\
&\quad +
\underbrace{\sum_{s=\pm}\Big[\loss_{\text{shape},s,\theta} + \lambda_{\text{int}}\,\loss_{\text{int},s,\theta}\Big]}_{\text{unconditional NCV}}
+ \loss_{\text{NIS},\phi}
+ \lambda_{\text{neg}}\,\loss_{\text{neg}}\; .
\label{eq:loss_nested_total}
\end{align}
%

\subsection{Performance metrics}
\label{sec:performance_metrics}

We benchmark our NCVs on a set of performance metrics. They are evaluated on the inference weights of Eq.\eqref{eq:cv_mc}: $w_\phi=f/g_\phi$ for pure NIS or $w_{\text{res},\phi,\theta}$ with NCV. The estimator always splits into a residual and a trivially sampled NCV contribution,
\begin{align}
\sigma
=
\sigma_{\text{res},\theta}+ C_\theta
=
\sigma_{\text{res},\theta}^+ - \sigma_{\text{res},\theta}^- + C_\theta\;.
\label{eq:sigma_split}
\end{align}
We quantify the cost of the residual with:
\begin{itemize}
\item Estimated cross section and its positive and negative parts,
\begin{align}
    \sigma &= \left\langle w \right\rangle = \frac{1}{N}\sum_i w_i\notag\\
    \mand \quad
    \sigma^\pm &=\left\langle \max\!\left( 0,\, \pm w \right)\right\rangle
    \qquad \Rightarrow \qquad
    \sigma =\sigma^+ - \sigma^-\;,
    \qquad \sigma^\text{abs}=\sigma^+ + \sigma^-\;.
\end{align}
For the residual weight $w_{\text{res},\phi,\theta}$ these are the residual contributions $\sigma_{\text{res},\theta}^\pm$, while for the plain weight $w_\phi$ without an NCV the contribution is the full cross section $\sigma$.

\item Sign floor on the relative variance.
Even a perfect importance sampler cannot remove the fluctuation from the sign of the weights. After flattening the absolute value of the weights, $|w|=C$, their values are $+C$ with probability $\rho$ and $-C$ with probability $1-\rho$,
\begin{align}
\rho = \frac{\sigma^+}{\sigma^+ + \sigma^-}
\qquad \Rightarrow \qquad
\sigma = C(2\rho-1)
\qquad \mand \qquad
\text{Var}(w) = 4C^2\rho(1-\rho)\; .
\end{align}
The relative variance is bounded from below,
\begin{align}
\text{RV} \ge \text{RV}_{\min}
=
\frac{\text{Var}(w)}{\sigma^2}
=
\frac{4\rho(1-\rho)}{(2\rho-1)^2}\;.
\label{eq:rv_floor}
\end{align}
This floor is non-zero whenever there are positive and negative weights.

\item Relative variance of a set of weights,
\begin{align}
    \text{RV}[w]
    =
    \frac{\langle w^2 \rangle - \langle w \rangle^2}
         {\langle w \rangle^2} \; .
    \label{eq:rv_def}
\end{align}
The self-normalized relative variance of the residual alone can be changed arbitrarily by shifting a constant between the sampled and integrated terms, since $\sigma=C_\theta+\langle w - C_\theta\rangle$ with $\text{Var}(w)=\text{Var}(w-C_\theta)$.
For each sampled contribution $w_a$ we therefore quote its contribution to the relative variance of the full target,
\begin{align}
    \text{RV}_{\sigma}[w_a]
    =
    \frac{\langle w_{a}^2 \rangle - \langle w_{a} \rangle^2}
         {\sigma^2}
    \qquad \Rightarrow \qquad
    \frac{\text{Var}(\hat\sigma)}{\sigma^2}
    =
    \sum_a \frac{1}{N_a}\,\text{RV}_{\sigma}[w_a] \; .
    \label{eq:rv_sigma_def}
\end{align}
Contributions known analytically or estimated at negligible cost, such as $C_\theta$ in Eq.\eqref{eq:sigma_split}, are assigned negligible statistical uncertainty.

\item Reduction in effective statistics, or Kish factor, due to weight fluctuations and cancellations,
\begin{align}
    \frac{N_\text{eff}}{N}
    =
    \frac{1}{N}
    \frac{\left(\sum_i w_i\right)^2}{\sum_i w_i^2}
    =
    \frac{\langle w \rangle^2}{\langle w^2 \rangle}
    =
    \frac{1}{\text{RV}[w] + 1} \; .
    \label{eq:kish_def}
\end{align}
For the residual weight, this characterizes the residual sample itself, while $\text{RV}_{\sigma}[w_{\text{res},\phi,\theta}]$ measures its contribution to the relative variance of the full cross section.
\end{itemize}

\subsection{Toy examples}
\label{sec:method_toy}

Before turning to physics applications, we illustrate the single and nested NCVs on controlled integrands. We use a two-dimensional toy version of Eq.\eqref{eq:nested_split} on the unit square,
\begin{align}
\sigma
= \int_0^1 \dd x_B\,\dd x_R\; f(x_B,x_R)
\qquad \mwith \qquad
f(x_B,x_R)= b(x_B)+r(x_B,x_R)\;.
\label{eq:signtoy_def}
\end{align}
We compare the single joint NCV with one control variate $c_\theta$ with the nested NCV, combining a conditional NCV $c_\eta(x_R|x_B)$ and an unconditional NCV $c_\theta(x_B)$. The first toy example comes with a sign change in $x_B$ and another one in $x_R$,
\begin{align}
\text{Toy-I}: \qquad
 b(x_B) &= x_B(1-x_B) + x_B(1-x_B)\,\Bigg[ 2 - 2 \Big[1+5 \ln(1+x_B(1-x_B)) \Big]\Bigg] \,\sin(10 x_B) \notag \\
r(x_B,x_R) &= x_B(1-x_B) \; \Bigg[- (1+x_R) + 1.6 -\frac{1}{2} \sin(15 x_R) \Bigg] \eqperiod
\end{align}
The sine terms determine the negative-weight fractions in the two variables. Because the integrand changes sign in $x_B$ as well as in $x_R$, the negative contribution persists after integrating over $x_R$. It is smooth, with a wide and easily populated negative region contributing $\sigma^-/\sigma=0.105$ to the cross section.

The negativity encountered in a typical NLO calculation is narrow, deep, and non-trivially shaped. Our second toy example models such a structure,
\begin{align}
\text{Toy-II}: \qquad
b(x_B) &= 6\,x_B(1-x_B)
\notag \\
r(x_B,x_R) &= b(x_B)\,
\Big[\, 0.4 - 40\, e^{-\left( (x_R - 0.1 - 0.6\,x_B)/0.003\right)^2} \Big]\eqperiod
\label{eq:def_toy_hard}
\end{align}
The combined integrand is positive almost everywhere, with a narrow negative slice of width $0.003$ in $x_R$ contributing $\sigma^-/\sigma = 0.165$. The center of this slice follows the correlation $x_R=0.1+0.6\,x_B$, so a factorized density cannot trace it. It is an analog of a subtraction-induced dip whose position in the radiation phase space depends on the Born configuration.

\subsubsection*{Numerical results}
\label{sec:method_toy_num}

\begin{figure}[b!]
    \includegraphics[width=0.99\linewidth]{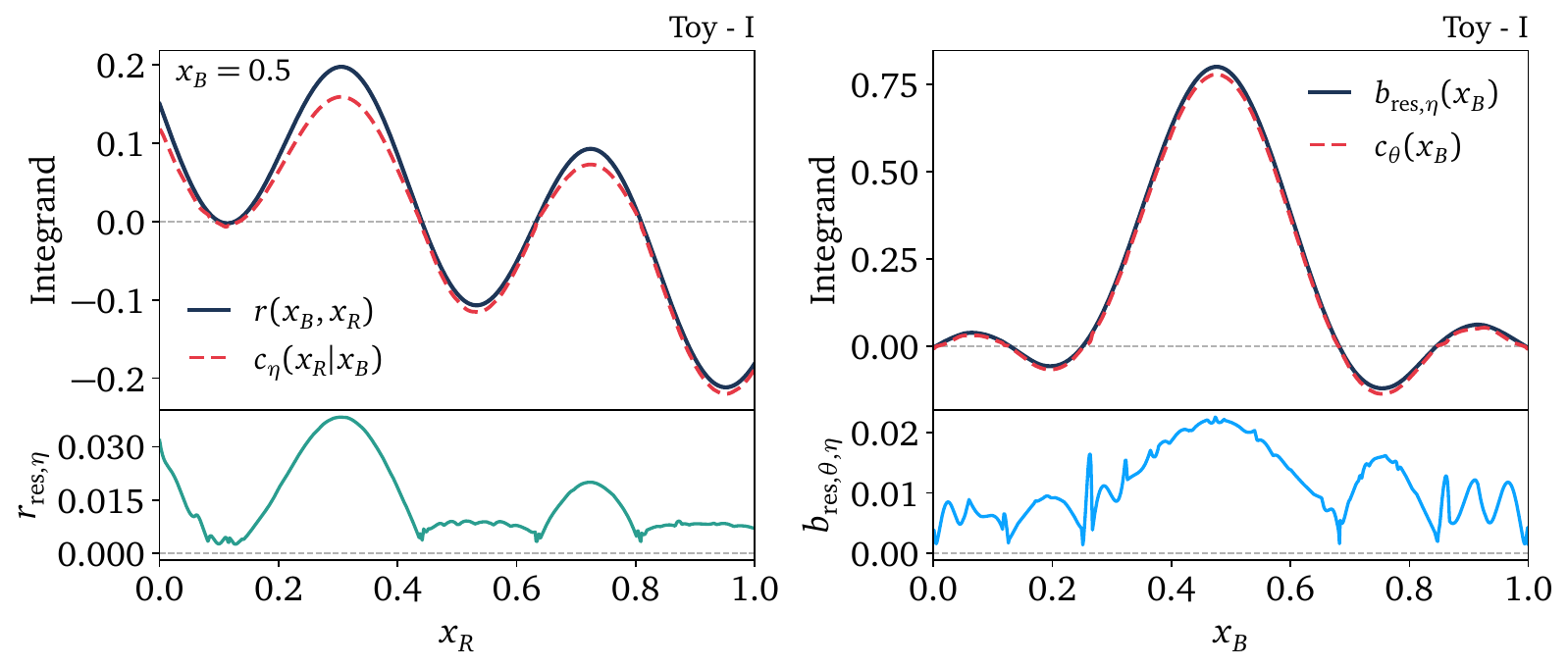}
    \caption{Nested NCV for Toy-I. Left: real-emission integrand $r$ at $x_B=0.5$ together with the conditional NCV, and the residual underneath. Right: the same at the unconditional level as a function of $x_B$.}
    \label{fig:toy1_visualize}
\end{figure}

In the left panel of Figure~\ref{fig:toy1_visualize}, we show $r(x_B=0.5,x_R)$ for Toy-I, with a sign-changing oscillation in two dimensions. The right panel shows how the NCV learns the negative $b_{\text{res},\eta}(x_B)$ and leaves us with a positive integrand. We compare the numerical evaluations
\begin{enumerate}
    \item Uniform sampling
    \item \madnis sampling
    \item single NCV (1-NCV + \madnis)
    \item nested NCV (2-NCV + \madnis)
\end{enumerate}
Following Section~\ref{sec:method_loss}, we normalize the shape and integral losses by the contribution of their targets. For the NIS likelihood and softplus normalizations, we choose the contribution of the negative component,
\begin{alignat}{9}
Z_\text{NIS} &= Z_\text{neg} = C_{-,\theta}\big|_{\text{no-grad}}
&&\qqqquad \text{(1-NCV)}
\notag \\
Z_\text{NIS} &= Z_{\text{neg},b} = C_{-,\theta}\big|_{\text{no-grad}}
\qquad \mand \qquad
Z_{\text{neg},r} = Z_{-,A}
&&\qqqquad \text{(2-NCV)} \;,
\label{eq:toy_norm_nested}
\end{alignat}
where $Z_{-,A}$ is the conditional contribution of Eq.\eqref{eq:nested_normalizations}. The penalty is measured in units of the structure that has to be lifted rather than the full integrand. Since this normalization is degenerate with $\beta_\text{max}$, it is a choice of units, and the values in the hyperparameter Table~\ref{tab:toy_hyperparameters} refer to it.

\begin{table}[t]
\centering
\setlength{\tabcolsep}{6pt}
\renewcommand{\arraystretch}{1.2}
\begin{small}
\begin{tabular}{ll|S[table-format=+1.3]S[table-format=1.3]cS[table-format=1.2e2]S[table-format=4.1]}
\toprule
Setup & Method
& {$C_\theta/\sigma$}
& {$\sigma_\text{res}^+/\sigma$}
& {$\sigma_\text{res}^-/\sigma$}
& {RV$_\sigma\,(\text{res})$}
& {RV$_{\madnis}$/RV} \\
\midrule
\multirow{4}{*}{Toy-I}
& Uniform                  & {--}   & 1.105 & $1.05\cdot10^{-1}$ & 2.47e0  & 0.2    \\
& \madnis                  & {--}   & 1.105 & $1.05\cdot10^{-1}$ & 4.78e-1 & 1.0    \\
& 1-NCV $+$ \madnis  & +0.922 & 0.078 & $8.76\cdot10^{-6}$ & 3.86e-4 & 1241.0 \\
& 2-NCV $+$ \madnis  & +0.895 & 0.105 & $5.16\cdot10^{-7}$ & 1.38e-4 & 3464.0 \\
\midrule
\multirow{4}{*}{Toy-II}
& Uniform                  & {--}   & 1.165 & $1.65\cdot10^{-1}$ & 5.22e0  & 0.1    \\
& \madnis                  & {--}   & 1.165 & $1.65\cdot10^{-1}$ & 7.73e-1 & 1.0    \\
& 1-NCV $+$ \madnis  & +0.208 & 0.812 & $2.05\cdot10^{-2}$ & 9.64e-2 & 8.0    \\
& 2-NCV $+$ \madnis  & +0.464 & 0.536 & $2.18\cdot10^{-5}$ & 1.32e-2 & 58.0   \\
\bottomrule
\end{tabular}
\end{small}
\caption{Performance metrics for the two toy models, evaluated on 1M points.
The first three columns show the trivial NCV contribution and the sampled residual adding to $\sigma$.
We do not optimize the hyperparameters, so the numbers indicate the size of the effect rather than the optimum.}
\label{tab:metrics-toy-cv}
\end{table}

In Table~\ref{tab:metrics-toy-cv} we show that the two toy setups behave very differently. Importance sampling alone leaves the negative-weight fraction untouched. The NCV removes the sign and moves most of the integral from the sampled residual into the NCV contribution $C_\theta$. For Toy-I, the positive residual drops to the point where $90\%$ of the integral comes from $C_\theta$. The negative contribution is reduced by four to five orders of magnitude. Both NCVs remove the sign equally well and reduce the relative variance by factors of $1241$ and $3464$ over \madnis.

For Toy-II, the single NCV leaves $\sigma_\text{res}^-/\sigma = 2.1\cdot10^{-2}$ and gains only a factor $8$ over \madnis, while the nested construction properly removes negative regions and gains a factor $58$. The reason is the shape of the negative slice, which the single NCV has to resolve as a thin diagonal structure in the $(x_B,x_R)$ space, whereas the conditional flow $p_{-,\eta}(x_R|x_B)$ only has to place a narrow peak at the correct $x_R$ given $x_B$.

The remaining gap to a perfect result is determined by the sign structure of the integrand itself. Using the \madnis values of Table~\ref{tab:metrics-toy-cv} we estimate
\begin{align}
\text{Toy-I}: \quad
\rho &= \frac{1.105}{1.105+0.105} = 0.913
&&\Rightarrow \quad
\text{RV}_{\min} = 0.466
\quad \text{vs.} \quad
\text{RV}_{\madnis} = 0.478
\notag \\
\text{Toy-II}: \quad
\rho &= \frac{1.165}{1.165+0.165} = 0.876
&&\Rightarrow \quad
\text{RV}_{\min} = 0.765
\quad \text{vs.} \quad
\text{RV}_{\madnis} = 0.773\;.
\end{align}
In both cases \madnis comes within $3\%$ of this floor, so its remaining variance is dominated by the intrinsic sign cancellation rather than by non-optimal sampling. The floor can only be lowered by modifying the integrand, in our case through the NCV.

\clearpage
\section{NCV range reduction at LO}
\label{sec:lo}

We first apply the NCV to a LO cross section with a positive integrand. It compresses the range of event weights and, combined with \madnis~\cite{Heimel:2022wyj,Heimel:2023ngj,Heimel:2024wph} and \madspace~\cite{Heimel:2026hgp}, improves the unweighting efficiency. Our training goal is a constant, small, and positive integrand.

\subsection{Phase-space integral}
\label{sec:lo_intro}

The LO cross section for the production of $n$ particles is
\begin{align}
    \sigma 
    = \int \dd \Phi_n\,B(\Phi_n)
    \qqquad\mwith \qqquad 
    \Phi_n \in \mathbb{R}^{3n-4} \; .
    \label{eq:psinteg_lo}
\end{align}
Detector acceptance and experimental selection criteria define a fiducial cross section
\begin{align}
\sigma_\text{fid}
=
\int \dd\Phi_n\; \jcuts (\Phi_n)\,B(\Phi_n)\,
\qquad \mwith \qquad
\jcuts (\Phi_n)
=
\begin{cases}
1 & \text{accepted} \\
0 & \text{rejected} \; .
\end{cases}
\label{eq:psinteg_lo_fid}
\end{align}
To integrate it, we parameterize the phase space using an analytic \madspace mapping,
\begin{align}
x_B\in [0,1]^{3n-4}
\quad \xleftrightarrow{\quad\text{\madspace $g_B(\Phi)$}\quad} \quad
\Phi_n \; ,
\label{eq:born_mapping_lo}
\end{align}
with the Jacobian 
\begin{align}
g_B(\Phi_n)
&= \left|
\frac{\partial x_B(\Phi_n)}{\partial \Phi_n}
\right|
\qqquad \mand \qqquad
\int \dd\Phi_n\, g_B(\Phi_n)=1\;.
\label{eq:born_density_lo}
\end{align}
This allows us to write
\begin{align}
\sigma_\text{fid}
&=
\int \dd x_B\;\jcuts (\Phi_n(x_B))\,
\frac{B(\Phi_n(x_B))}{g_B(\Phi_n(x_B))} \notag \\
&\equiv
\int \dd x_B\;\jcuts (x_B)\,\hat{B}(x_B)\; ,
\label{eq:lo_unit_hypercube}
\end{align}
such that $\hat{B}$ is the remapped Born contribution onto the unit hypercube. We employ \madnis,
\begin{align}
r \in[0,1]^{3n-4}
\quad \xleftrightarrow{\quad\text{\madnis $g_\phi(x_B)$}\quad} \quad 
x_B
\quad \xleftrightarrow{\quad\text{\madspace $g_B(\Phi)$}\quad} \quad
\Phi_n\; .
\label{eq:born_mapping_chain}
\end{align}
with a learned Jacobian
\begin{align}
g_\phi(x_B)
&=
\left|
\frac{\partial r(x_B)}{\partial x_B}
\right|
\qqquad\mand \qqquad
\int \dd x_B\, g_\phi(x_B)=1\; .
\label{eq:madnis_density_x}
\end{align}
The Monte Carlo estimator for the fiducial cross section becomes
\begin{align}
\sigma_\text{fid}
&=\left\langle \jcuts (x_B)\,\frac{\hat{B}(x_B)}{g_\phi(x_B)}\right\rangle_{x_B\sim g_\phi(x_B)} \; .
\label{eq:lo_madnis_estimator}
\end{align}
A persistent problem, even for a learned $g_\phi(x_B)$, is underestimated tails with $g_\phi(x_B)\to0$ for $\hat{B}(x_B)\ne 0$, such that large local weights reduce the global unweighting efficiency. 

\subsection{NCV sampling}
\label{sec:lo_sample}

We employ the NCV introduced in Eq.\eqref{eq:ps_cv0} to minimize the variance and improve the unweighting efficiency. Since $\hat{B}(x_B)\ge0$ the negative component vanishes and the NCV reduces to
\begin{align}
\sigma_\text{fid}
&= \int \dd x_B\,
\jcuts (x_B) \; 
\Big[
    \hat{B}(x_B)
    -
    C_{+,\theta}\,p_{+,\theta}(x_B)
\Big] + 
C_{+,\theta}\,\int \dd x_B\,\jcuts (x_B)\, p_{+,\theta}(x_B) \; .
\label{eq:ps_cv}
\end{align}
This turns into the Monte Carlo estimate
\begin{align}
\sigma_\text{fid}
&=
\left\langle \jcuts (x_B)
\frac{\hat{B}(x_B) -C_{+,\theta}\,p_{+,\theta}(x_B)}{g_\phi(x_B)} \right\rangle_{x_B\sim g_\phi}
+ C_{+,\theta}\XLangle \jcuts (x_B)\XRangle_{x_B\sim p_{+,\theta}}
\notag\\
&=
\XLangle w_{\text{res},\phi,\theta}(x_B)\XRangle_{x_B\sim g_\phi}
+ \XLangle w_{\text{NCV}, \theta}(x_B)\XRangle_{x_B\sim p_{+,\theta}}
\notag\\
&= \sigma_{\text{res} , \theta} + \sigma_{\text{NCV},\theta}
\qquad \mwith \qquad
\sigma_{\text{NCV},\theta} = C_{+,\theta}\,\varepsilon_\text{cuts}^{p}
\quad \mand \quad
\varepsilon_\text{cuts}^{p} = \big\langle \jcuts (x_B)\big\rangle_{x_B\sim p_{+,\theta}}\;.
\label{eq:ps_cv2}
\end{align}
The NCV contribution $\sigma_{\text{NCV},\theta}$
only requires the evaluation of the cut function on samples from $p_{+,\theta}$, without expensive matrix element. We estimate $\varepsilon_\text{cuts}^{p}$ such that its statistical uncertainty is negligible and the variance of $\sigma_\text{fid}$ is dominated by $\sigma_{\text{res},\theta}$.

Phase-space points not passing cuts do not require a matrix-element evaluation, so we quote unweighting efficiencies for points entering the expensive event generation. They differ from the relative variance of the Monte Carlo estimator and lead to the residual 
\begin{align}
\epsilon^\text{res}_\text{uw}
    =
    \frac{\left\langle w_{\text{res},\phi,\theta}(x) \right\rangle_{w\ne 0}}
         {|w_\text{res}|_\text{max}} \;,
\label{eq:uw_eff_res}
\end{align}
where $|w|_\text{max}$ is typically chosen to allow for a small number of over-weight events. The unweighting efficiency of the combined event sample is then
\begin{align}
\epsilon_\text{uw}^\text{tot}
=
\frac{\sigma_{\text{res},\theta}^{\text{abs}}}
{\sigma_{\text{res},\theta}^{\text{abs}} + \sigma_{\text{NCV},\theta}}\,\epsilon^\text{res}_\text{uw}
+
\frac{\sigma_{\text{NCV},\theta}}
{\sigma_{\text{res},\theta}^{\text{abs}} + \sigma_{\text{NCV},\theta}}
\qquad \mwith \quad
\sigma_{\text{res},\theta}^{\text{abs}}
= \XLangle \big|w_{\text{res},\phi,\theta}(x_B)\big|\XRangle_{x_B\sim g_\phi}\; .
\label{eq:uw_eff_cv}
\end{align}
The smaller $\sigma_{\text{res},\theta}^{\text{abs}}$, the larger the event fraction that can be produced with the cheap NCV contribution. We aim to maximize $C_{+,\theta}$ while keeping the integrand positive. 

To compute $\sigma_{\text{res},\theta}$ we combine NCV and NIS and train three networks:
\begin{itemize} 
\item NCV flow $p_{+,\theta}(x_B)$;
\item scalar normalization $C_{+,\theta}$; and 
\item standard NIS flow $g_\phi(x_B)$.
\end{itemize}
We want to absorb as much of the cross section as possible into the trivial integral, \ie make $C_{+,\theta}$ large while keeping $\hat{B}(x_B) - C_{+,\theta}\,p_{+,\theta}(x_B)$ small, constant, and positive. 

We use the single-NCV training of Section~\ref{sec:method_loss} for our positive, cut-restricted integrand~\cite{neuralcontrol}. The positivity barrier serves as a safeguard against overshoot. All networks are trained on samples following
\begin{align}
q(x_B)
&=
\frac{p_{+,\theta}(x_B)+g_\phi(x_B)}{2}\Bigg|_{\text{no-grad}}\;,
\end{align}
which defines the target weight for the NCV shape and the residual weight for the sampler, in direct analogy to Section~\ref{sec:method_loss} but now including the cut function,
\begin{align}
w_+(x_B) &= \jcuts (x_B)\frac{\hat{B}(x_B)}{q(x_B)}
\qquad \mand \qquad
w_\text{NIS}(x_B) =
\jcuts (x_B)\frac{\hat{B}(x_B)-C_{+,\theta}\,p_{+,\theta}(x_B)\big|_{\text{no-grad}}}{q(x_B)}\; .
\label{eq:def_weights_LO}
\end{align}
At LO only the positive component is active, so all normalizations are set by the cut-weighted NCV contribution,
\begin{align}
Z_+ = Z_\text{NIS} = Z_\text{neg} \equiv Z_C
= \sigma_{\text{NCV},\theta}\big|_{\text{no-grad}}
= C_{+,\theta}\,\varepsilon_\text{cuts}^{p}\big|_{\text{no-grad}} \;.
\label{eq:norm_cuts}
\end{align}
The shape loss for $p_{+,\theta}$ and the NIS loss for $g_\phi$ take the form of Eqs.\eqref{eq:loss_shape_single_ncv} and~\eqref{eq:loss_is_single_ncv} with the normalizations above. Because the NCV enters the fiducial cross section only through $\sigma_{\text{NCV},\theta}=C_{+,\theta}\,\varepsilon_\text{cuts}^{p}$, the integral loss targets this combination,
\begin{align}
\loss_{\text{int},\theta}
=
\frac{1}{Z_C^2}
\left[
C_{+,\theta}\,\varepsilon_\text{cuts}^{p}
-
\big\langle
w_+(x_B)
\big\rangle_{x_B\sim q}
\right]^2\;.
\label{eq:loss_int_x}
\end{align}
The positivity barrier acts only on accepted events, \ie with the cut function as a prefactor,
\begin{align}
\loss_{\text{neg},\theta}
=
\XXXLangle
\jcuts (x_B)\;
\Psi \!\left(\frac{\hat{B}(x_B)-C_{+,\theta}\,p_{+,\theta}(x_B)}{Z_C\,q(x_B)}\right)
\XXXRangle_{x_B\sim q}\; .
\label{eq:loss_neg_x}
\end{align}
A cut penalty discourages $p_{+,\theta}$ from sampling outside the accepted region,
\begin{align}
\loss_{\text{cut},\theta}
=
\XXXLangle
(1-\jcuts (x_B))\;
\Psi \!\left(\frac{\hat{B}(x_B)-C_{+,\theta}\,p_{+,\theta}(x_B)}{Z_C\,q(x_B)}\right)
\XXXRangle_{x_B\sim q}\; ,
\label{eq:loss_cut_x}
\end{align}
so that the full loss reads
\begin{align}
\loss =
\loss_{\text{shape},+,\theta}
+ \lambda_\text{int}\,\loss_{\text{int},+,\theta}
+ \loss_{\text{NIS},\phi}
+ \lambda_\text{neg}\,\loss_{\text{neg},\theta}
+ \lambda_\text{cut}\,\loss_{\text{cut},\theta}\; .
\label{eq:loss_alch_x}
\end{align}
The negative-weight and cut penalties could be merged, but cut events are computationally much less burdensome than negative events, so we control them separately.

\subsection{Top pair production with gluons}
\label{sec:lo_phys}

As a realistic benchmark, we look at the LO process
\begin{align}
 \Pg \Pg \to \Pt \Ptbar \Pg \Pg \; ,
\end{align}
with stable tops at 13~TeV center-of-mass energy, using the default NNPDF2.3 LO PDF set~\cite{Ball:2012cx} in \mg~\cite{Alwall:2014hca}. The matrix element is generated using the \cudacpp plugin~\cite{Hagebock:2025jyk}. We use a single-channel \madspace mapping with a pure $t$-channel topology and compare three setups:
\begin{enumerate}
    \item NIS-NCV allowing for negative integrands, \ie $\lambda_\text{neg}=0$;
    \item NIS-NCV penalizing negative integrands, \ie $\lambda_\text{neg}=10$;
    \item a single-channel \madnis setup with only NIS.
\end{enumerate}
We provide the training hyperparameters in Table~\ref{tab:lo_hyperparameters}. For validation, we generate 10M unweighted events for the same process using \madspace. We follow Eq.\eqref{eq:uw_eff_cv} and optimize $C_{+,\theta} p_{+,\theta}$ and $g_\phi$ so that the integrand of $\sigma_{\text{res} , \theta}$ has low variance and, ideally, zero negative points. Under that condition, we make $C_{+,\theta}$ as large as possible.

\begin{figure}[t!]
    \includegraphics[width=0.49\linewidth,page=1]{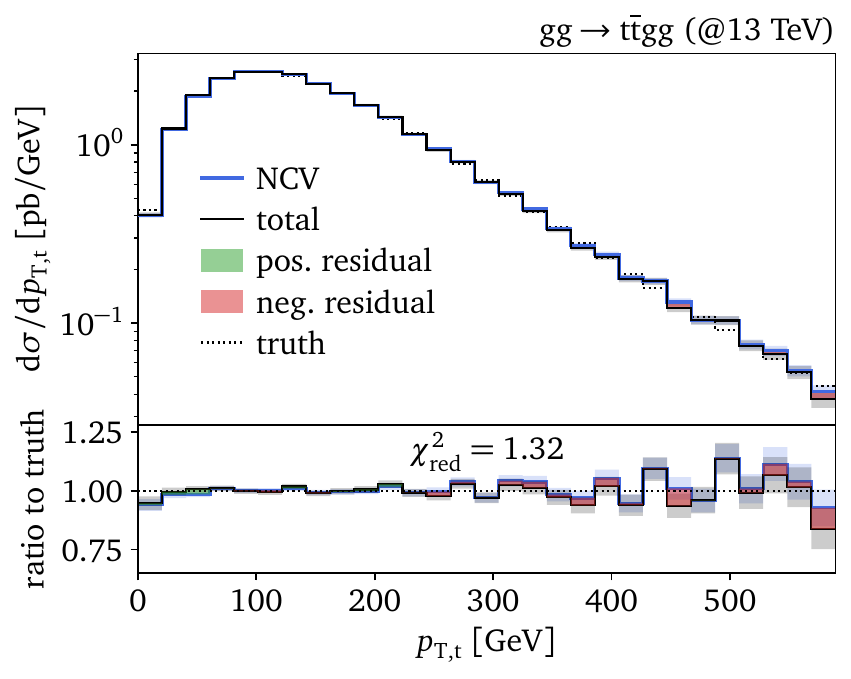}
    \includegraphics[width=0.49\linewidth,page=1]{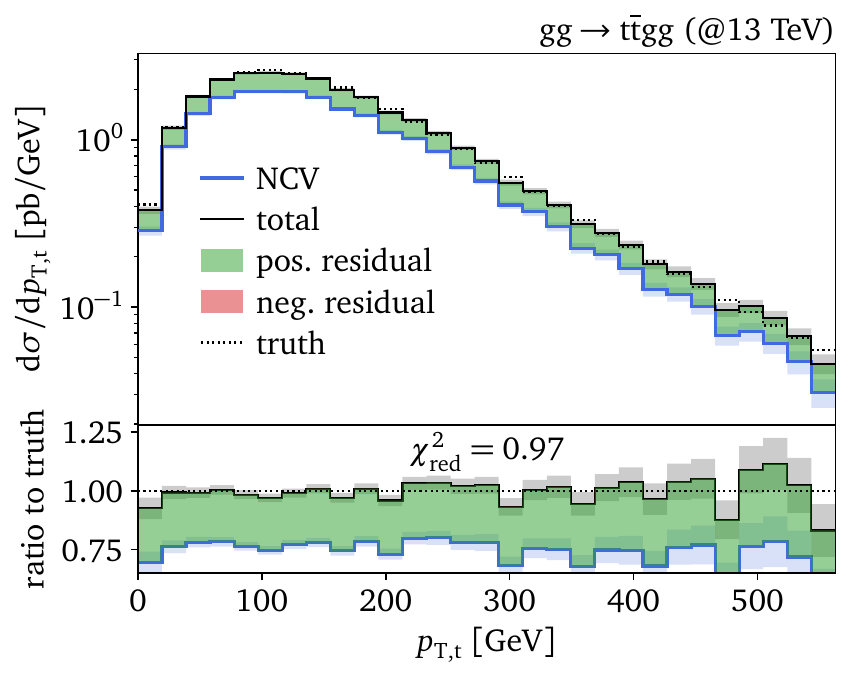}
    \caption{Top transverse momentum distributions $p_{\mathrm{T},\mathrm{t}}$ for $\Pt \Ptbar \Pg \Pg$ production using an NCV trained without (left) and with (right) $\loss_{\text{neg},\theta}$ to penalize negative weights.}
    \label{fig:ttgg}
\end{figure}

We show the $p_{\text{T},\Pt}$ distributions in Figure~\ref{fig:ttgg}. To the left, we see that $C_{+,\theta}$ constitutes a large fraction of the total cross section but overshoots slightly in some bins. This leads to a negative residual contribution. For the right panel, we penalize negative integrands and confirm that only a small positive residual contribution is needed to correct all bins.

\begin{table}[b!]
\setlength{\tabcolsep}{8pt}
\centering
{\renewcommand{\arraystretch}{1.2}
\begin{small}
\begin{tabular}{l|c c c}
    \toprule
    Metric & $\lambda_\text{neg} = 10$ & $\lambda_\text{neg} = 0$ & only NIS \\
    \midrule
    $\sigma_{\text{NCV}} / \sigma$ & 0.767 & 1.001 &  {--}\\
    $\sigma^+_\text{res} / \sigma$ & 0.238 & 0.137 & 1.000\\
    $\sigma^-_\text{res} / \sigma$ & 0.005 & 0.137 & {--} \\
    \midrule
    $N_\text{eff}/N \text{ (NCV) } = \varepsilon_\text{cuts}^{p}$ &  0.189 &  0.948  & {--}\\
    $N_\text{eff}/N$ (res) &  0.408 &  0.000  & 0.725\\
    $\text{RV}_{\sigma}\,(\text{res})$ & 0.078 & 0.196 & 0.379 \\
    \midrule
    $\epsilon^\text{res}_\text{uw}$ & 0.149 & 0.002 & 
    0.294\\
    $\epsilon^\text{tot}_\text{uw}$ & 0.796 & 0.786 & 
    0.294\\
    $f^\text{uw}_\text{gain}$ &  2.708 &  2.673  & 1.000\\
    \bottomrule
\end{tabular}
\end{small}}
\caption{Metrics for the $\Pg \Pg \to \Pt \Ptbar \Pg \Pg$ cross section with and without penalizing negative weights. All unweighting efficiencies are given allowing over-weights to contribute $1\%$ of the total cross section.}
\label{tab:ttgg}
\end{table}

In Table~\ref{tab:ttgg} we compare the performance metrics from Section~\ref{sec:performance_metrics}, evaluated for the two contributions in Eq.\eqref{eq:def_weights_LO}. In addition, we show the residual and total unweighting efficiencies given in Eqs.\eqref{eq:uw_eff_res} and~\eqref{eq:uw_eff_cv}, and the gain in effective unweighted statistics relative to pure NIS,
\begin{align}
        f^\text{uw}_\text{gain}
        =
        \frac{\epsilon^\text{tot}_\text{uw}}{\epsilon_\text{uw}^{\text{(NIS)}}}\;.
\end{align}
The NCV substantially reduces the size of the residual integral, which is then handled efficiently by NIS. With penalized negative weights, the residual contribution is almost entirely positive, resulting in a Kish factor $N_\text{eff}/N$ close to unity and a negligible loss in effective statistics. 

Compared to pure NIS, the unweighting efficiency clearly benefits from the NCV, with and without negative-weight penalty. Without penalty, negative residual events reduce the effective statistics of the final event sample. With penalty, the residual contribution is larger but almost entirely positive. Both cases have similar unweighting efficiencies $\epsilon^\text{tot}_\text{uw}$, \ie a similar number of matrix element evaluations. However, negative-weight events increase the computational cost in downstream simulation steps.

We find a gain in the unweighting efficiency of $f^\text{uw}_\text{gain}\simeq 2.7$, with and without the negative-weight penalty. This gain comes on top of the improvement of roughly an order of magnitude from \madnis over \mg.

This event-generation metric is different from the integration gain inferred from the relative variance. For $\lambda_\text{neg}=10$, the residual contribution to the relative variance of the fiducial cross section is $\text{RV}_{\sigma}(\text{res})\simeq0.078$, compared to $0.379$ for pure NIS. Assuming that $\varepsilon_\text{cuts}^{p}$ comes with negligible statistical uncertainty, this corresponds to a reduction in the integration variance by $0.379/0.078\simeq4.9$ relative to pure NIS. Thus, NCV and NIS yield complementary practical benefits even for relatively simple LO integrals.

\clearpage
\section{NCV sign removal at NLO}
\label{sec:nlo}

The next-to-leading order (NLO) cross section corresponding to Eq.\eqref{eq:psinteg_lo} has the form
\begin{align}
    \sigma_\text{NLO} 
    &= \int \dd \Phi_n\, \left[B(\Phi_n) + V(\Phi_n)\right] + \int \dd \Phi_{n+1}\; R(\Phi_{n+1})\; ,    
\label{eq:nlo_def}
\end{align}
where $V(\Phi_n)$ describes the virtual correction and $R(\Phi_{n+1})$ is the real emission.
Both contributions contain soft and collinear divergences, and only their sum is finite. To enable numerical cancellation, we use subtraction terms $S(\Phi_{n+1})$ that reproduce the singular behavior of the real emission. After integrating over the radiation phase space $\dd\Phi_R$ they cancel the virtual divergence~\cite{Catani:1996vz,Catani:2002hc,Frederix:2009yq},
\begin{align}
    \sigma_\text{NLO} 
    &= \int \dd \Phi_n\,\big[B(\Phi_n) 
    + \underbrace{V(\Phi_n) + I(\Phi_n)}_{\textstyle \equiv V_\text{fin}(\Phi_n)}\big] 
    + \int \dd \Phi_{n+1}\,\big[ \underbrace{R(\Phi_{n+1}) - S(\Phi_{n+1})}_{\textstyle \equiv R_\text{fin}(\Phi_{n+1})}\big]\notag\\
&\mwith \qquad I(\Phi_n) = \int \dd \Phi_R\, S(\Phi_{n+1})\;.
\label{eq:subtraction_def}
\end{align}
Negative weights arise from two sources. First, $B(\Phi_n)+V(\Phi_n)$ can become negative because NLO truncates the squared amplitude after the interference term. This is a sign of large NLO corrections and no problem as long as the observable differential rate remains positive. Second, $R(\Phi_{n+1}) - S(\Phi_{n+1})$ can change sign, either because the subtraction term reproduces the real emission only in the singular limits and locally exceeds it elsewhere, or because the phase-space cuts on the re-mapped kinematics and on the $(n+1)$-body kinematics do not match. The latter can also be mitigated by an improved subtraction or measurement function.

For both sources of negative weights, we implement two NCVs from Section~\ref{sec:method}. A conditional NCV lifts the subtraction-induced fluctuations in the radiation variables, complementing the physics-motivated subtraction terms. An unconditional NCV lifts the residual sign changes in the Born-like kinematics.

\subsection{Soft and collinear subtraction}
\label{sec:nlo_tops}

\begin{figure}[t]
    \includegraphics[width=0.30\textwidth]{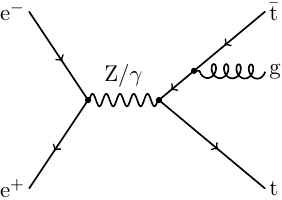}
    \hfill
    \includegraphics[width=0.30\textwidth]{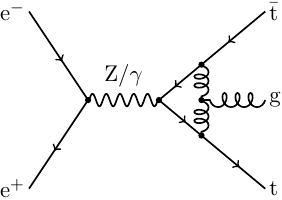}
    \hfill
    \includegraphics[width=0.30\textwidth]{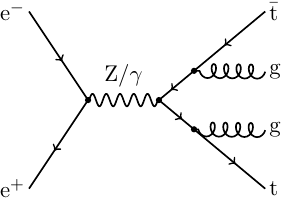}
    \caption{Representative Feynman diagrams for the Born, virtual, and real-emission contribution for the NLO predictions of the $\Pep \Pem \to \Pt \Ptbar \Pg$ process.}
    \label{fig:diagrams_4_jet}
\end{figure}

We consider the NLO QCD corrections to two processes
\begin{align}
\Pep \Pem &\to \Pt \Ptbar \Pg
\notag \\
\Pep \Pem &\to \Pq \Pqbar \Pg\;.
\label{eq:born_nlo}
\end{align}
For the first process we show example Feynman diagrams in Figure~\ref{fig:diagrams_4_jet}. For our technical introduction, we follow top pair production with a gluon. The massless three-jet processes can be treated analogously, as described in Appendix~D of Ref.~\cite{Catani:1996vz}.

To regularize the real-emission corrections, we employ Catani--Seymour (CS) dipole subtraction~\cite{Catani:1996vz,Catani:2002hc}, where the local subtraction term is a sum over dipoles,
\begin{align}
S(\Phi_{n+1})
= \sum_{(ij,k)}
\mathcal D_{ij,k}(\Phi_{n+1}) \; .
\label{eq:dipoles}
\end{align}
Each dipole corresponds to a singular configuration in which parton $j$ is soft or collinear to the emitter $i$, with parton $k$ acting as the spectator. This universal dipole subtraction relies solely on the color and spin structure. Each dipole factorizes into the Born matrix element evaluated for dipole-mapped $n$-body kinematics and a universal splitting kernel,
\begin{align}
\mathcal{D}_{ij,k}
=
-\frac{1}{(p_i+p_j)^2-m^2_{ij}}
\left\langle
\mathcal{M}_n(\tilde \Phi_n)
\left|
\frac{T_k \!\cdot\! T_{ij}}{T_{ij}^2}\,
V_{ij,k}
\right|
\mathcal{M}_n(\tilde \Phi_n)
\right\rangle \; .
\label{eq:cs_dipole}
\end{align}
Here, $V_{ij,k}$ denotes the CS splitting kernel, $T_{ij}$ and $T_k$ are color charge operators associated with the emitter and the spectator, and $\tilde \Phi_n$ denotes the re-mapped momenta ensuring momentum conservation and on-shell conditions in the reduced Born-like configuration. Integrating over the radiation phase space gives us
\begin{align}
I(\Phi_n)
&= \int \dd\Phi_R\; S\left(\Phi_{n+1}(\Phi_n,\Phi_R)\right) \notag \\
&=\left\langle
\mathcal{M}_n(\Phi_n)
\left|
I(\epsilon)
\right|
\mathcal{M}_n(\Phi_n)
\right\rangle\; ,
\label{eq:cs_integrated_def}
\end{align}
where the universal insertion operator $I(\epsilon)$ cancels the virtual singularities. It depends only on the color charges and kinematics of the external partons.

The acceptance cuts defining a fiducial cross section and the differential observables  are encoded in a measurement function $J$. For the NLO prediction to be finite, $J$ has to be infrared safe, \ie it has to become independent of an unresolved emission,
\begin{align}
J(\Phi_{n+1}) \;\xrightarrow{\quad\text{soft/collinear}\quad}\; J(\tilde\Phi_n)\,.
\label{jet_definition}
\end{align}
For the real emission it is evaluated on the $(n+1)$-body kinematics. 
For the subtraction term, it is evaluated on the respective dipole-mapped Born kinematics, ensuring the local cancellation of the singularities. Equation~\eqref{eq:nlo_def} then becomes
\begin{align}
\sigma_\text{NLO}
&=
\int \dd\Phi_n \;J(\Phi_n) \;\Big[B(\Phi_n)+V(\Phi_n)+I(\Phi_n)\Big] \notag \\
&+
\int \dd\Phi_{n+1} \; \left[ J(\Phi_{n+1}) \;R(\Phi_{n+1})
-\sum_{(ij,k)}\mathcal D_{ij,k}(\Phi_{n+1})\;
J\!\left(\tilde\Phi_n^{(ij,k)}(\Phi_{n+1})\right) \right]\;.
\label{eq:nlo_cs_with_measurement}
\end{align}
%

\subsubsection*{Dipole subtraction recap}

For top pair production with a gluon, the first of two real-emission processes is
\begin{align}
\Pep \Pem &\to \Pt(p_1)\,\Ptbar(p_2)\;\Pg(p_3)\,\Pg(p_4)\; ,
\label{eq:real_channels1}
\end{align}
with $(1,2,3,4)=(\Pt,\Ptbar,\Pg,\Pg)$ and the infrared singular limits
\begin{align}
p_3\to 0 \qqqquad p_4\to 0 \qqqquad p_3\parallel p_4\;.
\label{eq:ir_divergences}
\end{align}
Collinear limits involving tops, $p_{1,2}\parallel p_{3,4}$ are regularized by \Mt. The set of final--final dipoles treats each gluon once as unresolved, $i\in\{3,4\}$, and any colored leg $j \ne i$ as the emitter, and the remaining particle $k \ne i,j$ as the spectator. This gives ten dipoles
\begin{align}
(ij,k)\in\Big\{
&(31,2),(31,4),(32,1),(32,4),(34,1),\notag\\
&(43,2),(41,2),(41,3),(42,1),(42,3)
\Big\}\;.
\label{eq:dipoles_ttgg}
\end{align}
The sum of the dipoles reproduces the limits from Eq.\eqref{eq:ir_divergences}. Since the color algebra factorizes, we have to compute three types of dipoles
\begin{align}
    \mathcal{D}_{\Pg Q,Q}
    &= \frac{1}{2 p_\Pg\cdot p_Q}
    \left(1- \frac{C_A}{2C_F}\right)
    \braket{V_{\Pg Q,Q}}\,
    |\mathcal{M}_3(\tilde \Phi_n)|^2
    \notag \\
    \mathcal{D}_{\Pg Q,\Pg}
    &= \frac{1}{2 p_\Pg\cdot p_Q}
    \frac{C_A}{2C_F}
    \braket{V_{\Pg Q,\Pg}}\,
    |\mathcal{M}_3(\tilde \Phi_n)|^2
    \notag \\
    \mathcal{D}_{\Pg\Pg,Q}
    &= \frac{1}{2 p_3\cdot p_4}
    \frac{1}{2}\,
    \bra{\mathcal{M}_{3}^{\mu}(\tilde \Phi_n)}
    V_{\Pg\Pg,Q}^{ 
\mu\nu}
    \ket{\mathcal{M}_{3}^{\nu}(\tilde \Phi_n)}
    \qquad \text{with} \qquad
    Q\in\{\Pt,\Ptbar\} \; ,
    \label{spincorrelated_dipole}
\end{align}
The first two dipoles are straightforward to calculate using the spin-averaged splitting kernels $\braket{V_{ij,k}}$~\cite{Catani:2002hc}. For the gluon emitter we have to account for spin correlations, so the splitting kernel is a tensor in the spin space of the mapped parent gluon,
\begin{align}
    V_{ij,k}^{\mu\nu}
    \equiv
    \bra{\mu}V_{ij,k}\ket{\nu}.
\end{align}
By inserting a set of full helicity states~\cite{Frederix:2008hu} we project this tensor onto the physical gluon,
\begin{align}
    V_{\Pg\Pg,Q}(\lambda_b,\lambda_a)
    =
    \varepsilon_\mu(\lambda_b)\,
    V_{\Pg\Pg,Q}^{\mu\nu}\,
    \varepsilon_\nu^*(\lambda_a) \; .
\end{align}
The spin-correlated dipole can be evaluated as
\begin{align}
    \mathcal{D}_{\Pg\Pg,Q}
    =
    \frac{1}{2p_3\cdot p_4}
    \frac{1}{2}
    \sum_{\lambda_a,\lambda_b}
    \mathcal{M}_3^*(\lambda_b;\tilde\Phi_n)\,
    V_{\Pg\Pg,Q}(\lambda_b,\lambda_a)\,
    \mathcal{M}_3(\lambda_a;\tilde\Phi_n) \; .
\end{align}
The polarization vectors are evaluated using the HELAS routines in \mg. The diagonal terms, $\lambda_a=\lambda_b$, are obtained by multiplying the squared helicity amplitudes from \mg with the corresponding diagonal entries of the splitting matrix. The off-diagonal terms, $\lambda_a\neq\lambda_b$, require interference terms between reduced amplitudes with different helicities of the mapped parent gluon. These helicity amplitudes are stored before the helicity sum is performed.

The second real-emission process is
\begin{align}
\Pep \Pem &\to \Pt(p_1)\,\Ptbar(p_2)\;\Pq(p_3)\,\Pqbar(p_4)
\qquad \mwith \qquad q=\Pu,\Pd,\Ps,\Pc\; ,
\label{eq:real_channels2}
\end{align}
with $(1,2,3,4)=(\Pt,\Ptbar,\Pq,\Pqbar)$. We only encounter one collinear singularity 
\begin{align}
p_3\parallel p_4\; 
\end{align}
corresponding to the branching $\Pg\to \Pq\Pqbar$. We have to evaluate two dipoles:
\begin{align}
(ij,k)\in\Big\{ (34,1),\ (34,2)\Big\}\; ,
\label{eq:dipoles_ttqq} 
\end{align}
and their contribution can be obtained from Eq.\eqref{spincorrelated_dipole} by replacing $V_{\Pg\Pg,Q} \to V_{qq,Q}$. We use the fact that \mg returns the virtual contributions divided into
\begin{align}
    V = \bigg\{V_{\text{finite}}, V\!\left(\frac{1}{\epsilon}\right),V\!\left(\frac{1}{\epsilon^2}\right)\bigg \} \; .
    \label{eq:virtual}
\end{align}
We combine these finite terms with those from expanding Eq.(D.20) in Ref.~\cite{Catani:2002hc} in $\epsilon$.

\subsection{NCV sampling}
\label{sec:nlo_sampling}

The Born-like and real-emission phase spaces are generated using multi-channel techniques. The Born-like contribution to Eq.\eqref{eq:nlo_cs_with_measurement} is evaluated with the integrand
\begin{align}
f_n(\Phi_n)
=
\Big[ B(\Phi_n) + V(\Phi_n) + I(\Phi_n) \Big]\;
J(\Phi_n) \; .
\label{eq:fn_def}
\end{align}
The phase space is parameterized with a single-channel \madspace topology, Eq.\eqref{eq:born_mapping_lo}, with the Jacobian from Eq.\eqref{eq:born_density_lo}, leading to the Born-like contribution to $\sigma_\text{NLO}$ integrated over the unit hypercube
\begin{align}
\sigma_\text{NLO} 
\supset
\int \dd x_B\;
\frac{f_n(\Phi_n(x_B))}
     {g_B(\Phi_n(x_B))}
\equiv
\int \dd x_B\;
b(x_B) \; .
\label{eq:sigma_n_unit_hypercube}
\end{align}
%

\subsubsection*{Real emission}

For the real emission, we construct a multi-channel sampling density that targets the infrared-sensitive regions while keeping the CS subtraction untouched. The integrand in Eq.\eqref{eq:nlo_cs_with_measurement} is
\begin{align}
f_{n+1}(\Phi_{n+1})
&=
R(\Phi_{n+1})\,J(\Phi_{n+1})
-\sum_{(ij,k)}\mathcal D_{ij,k}(\Phi_{n+1})\;
J\!\left(\tilde\Phi_n^{(ij,k)}(\Phi_{n+1})\right)\,,
\label{eq:fn1_def}
\end{align}
In contrast to the LO case in Eq.\eqref{eq:ps_cv}, the cut function now affects three distinct kinematics: Born-like $\Phi_n$ in Eq.\eqref{eq:fn_def}, real emission $\Phi_{n+1}$, and dipole-mapped $\tilde\Phi_n^{(ij,k)}$ in Eq.\eqref{eq:fn1_def}. The fiducial selection enters training through the negative parts of the integrands. We build $\Phi_{n+1}$ by choosing a clustered--spectator pair out of six choices,
\begin{align}
(\widetilde{ij},\widetilde{k})
\in
\{
(\Pt,\Ptbar),(\Pt,\Pg),(\Ptbar,\Pt),
(\Ptbar,\Pg),(\Pg,\Pt),(\Pg,\Ptbar)
\} \; .
\label{eq:channel_set_six}
\end{align}
We define the corresponding $\Phi_{n+1}$ based on the inverse CS mapping and sampling of the radiation variables. Each configuration defines a channel $c$, and the full phase-space measure is given by the sum 
\begin{align}
    \dd\Phi_{n+1} = \sum_{c} \alpha_{c}\, \dd\Phi^{(c)}_{n+1}\,,
\end{align}
The CS mapping provides the clustered Born momenta $\tilde\Phi_n$ and the channel-dependent radiation variables $z_i$, $y_{ij,k}$ and $\varphi_i$,
\begin{align}
(\tilde{\Phi}_n,z_i,y_{ij,k},\varphi_i)\equiv(\tilde{\Phi}_n,\Phi^{(c)}_R)\; \xleftrightarrow[\quad\text{for each }(\widetilde{ij},\widetilde{k})\quad]{\quad\text{CS mapping}\quad} \; \Phi^{(c)}_{n+1}\equiv (p_1,\dots,p_n,p_{n+1})\; .
\end{align}
Each channel defines a Jacobian
\begin{align}
    g_{c}(\Phi^{(c)}_{n+1})=
   \left|
   \frac{\partial (\tilde{\Phi}_n,\Phi^{(c)}_R)}{\partial \Phi^{(c)}_{n+1}}
   \right|
   \qquad \mwith \qquad \int \dd\Phi^{(c)}_{n+1}\, g_{c}(\Phi^{(c)}_{n+1})=1
    \quad \mand \quad c\in\{(\widetilde{ij},\widetilde{k})\}\; ,
\end{align}
and the channels are combined as
\begin{align}
g_\text{CS}(\Phi^{(c)}_{n+1})
=
\sum_{c}
\alpha_c\; g_c(\Phi^{(c)}_{n+1})
\qquad \mwith \qquad
\sum_{c}\alpha_c=1\; .
\label{eq:global_density_real}
\end{align}
By default, we sample each channel with probability $\alpha_{c}=1/6$. This gives us for the real-emission contribution to $\sigma_\text{NLO}$ in Eq.\eqref{eq:nlo_cs_with_measurement}
\begin{align}
\sigma_\text{NLO}& \supset
\sum_{c} \alpha_c\int \dd \tilde{\Phi}_n\; \dd\Phi^{(c)}_R\;\frac{f_{n+1}\!\left(\Phi^{(c)}_{n+1}(\tilde{\Phi}_n,\Phi^{(c)}_R)\right)}{g_\text{CS}\left(\Phi^{(c)}_{n+1}(\tilde{\Phi}_n,\Phi^{(c)}_R)\right)}\notag\\
&\equiv\sum_{c} \alpha_c\int \dd \tilde{\Phi}_n\; \dd\Phi^{(c)}_R\;h_{n+1}\left(\tilde{\Phi}_n,\Phi^{(c)}_R\right)\;.
\label{eq:real_emission_ps_rad}
\end{align}
Finally, we have to decompose the real-emission phase space into Born-like and  radiation phase spaces, with the corresponding analytic mappings
\begin{align}
(x_B, x_R)\;
\xleftrightarrow{\quad\text{MadSpace}\quad}
\;(\tilde{\Phi}_n,\Phi^{(c)}_R)\;\xleftrightarrow{\quad\,\text{CS mapping}\,\quad}\; \Phi^{(c)}_{n+1} \; .
\label{eq:def_chan_cs}
\end{align}
The reduced Born-like phase space again follows Eq.~\eqref{eq:born_mapping_lo}. The radiation phase space requires three hypercube dimensions $x_R=(x_z,x_y,x_\varphi)$. For a massive final--final dipole $\mathcal{D}_{ij,k}$ their 
mapping to the radiation phase space is given by 
\begin{align}
y_{ij,k}(x_y)
&= y_-
 + \bigl(y_+-y_-\bigr)\, x_y^2 \notag \\
\varphi_i(x_\varphi)
&= 2\pi\,x_\varphi \notag\\
z_i(x_z;y)
&= z_-(y)
 + \left[ z_+(y)-z_-(y)\right]
   \sin^2\! \frac{\pi x_z}{2} \notag \\
g^{(c)}_R\!\left(\Phi^{(c)}_R\right)
&=
\frac{1}{
4\pi^2\,
\sqrt{(y_+-y_-)\,\bigl(y_{ij,k}-y_-\bigr)}\,
\sqrt{\bigl(z_i-z_-(y)\bigr)\,\bigl(z_+(y)-z_i\bigr)}
}\;.
\label{eq:rad_density}
\end{align}
The quadratic mapping in $y_{ij,k}$ and the trigonometric mapping in $z_i$ regularize the integrable singularities. After subtraction, the remaining dependence on $y_{ij,k}$ and $z_i$ typically exhibits square-root enhancements near the phase-space boundaries. This quadratic mapping does not smooth the collinear end point, but $z_i$ generates an integrable square-root scaling at both end points $z_i\to z_\pm$. As the sine transformation scales quadratically near $x_z\to 0,1$, it simultaneously smooths both end points. The integration boundaries are
\begin{align}
    y_- &= \frac{2\mu_i\mu_j}{1-\mu_i^2 -\mu_j^2-\mu_k^2} \qqquad  
    y_+ = 1-\frac{2\mu_k(1-\mu_k)}{1-\mu_i^2 -\mu_j^2-\mu_k^2} \notag\\
    z_\pm(y) &= \frac{2\mu_i^2+ (1-\mu_i^2-\mu_j^2-\mu_k^2)y_{ij,k}}{2[\mu_i^2 +\mu_j^2 +(1-\mu_i^2-\mu_j^2 -\mu_k^2)y_{ij,k}]} \left[ 1 \pm v_{ij,i}(y)\,v_{ij,k}(y) \right] \;,
\end{align}
where $v_{ij,k}$ ($v_{ij,i}$) is the relative velocity between $p_i + p_j$ and $p_k$ ($p_i$),
\begin{align}
v_{ij,k}(y) &= \frac{\sqrt{[2 \mu_k^2 + (1 - \mu_i^2 - \mu_j^2 - \mu_k^2)(1 - y_{ij,k})]^2 - 4 \mu_k^2}}{(1 - \mu_i^2 - \mu_j^2 - \mu_k^2)(1 - y_{ij,k})} \notag\\
v_{ij,i}(y) &= \frac{\sqrt{(1 - \mu_i^2 - \mu_j^2 - \mu_k^2)^2 \, y_{ij,k}^2 - 4\, \mu_i^2 \, \mu_j^2}}{(1 - \mu_i^2 - \mu_j^2 - \mu_k^2)\,y_{ij,k} + 2\mu_i^2}\;,
\end{align}
and 
\begin{align}
    \mu_i = \frac{m_i}{\sqrt{(p_i+p_j+p_k)^2}}\;.
\end{align}
Altogether, this gives us for the NLO cross section
\begin{align}
\sigma_\text{NLO}
&=\int \dd x_B\;b(x_B) 
+\sum_{c} \alpha_c\int \dd x_B\,\dd x_R\; \frac{h_{n+1}\Bigl(\tilde{\Phi}_n(x_B),\Phi_R^{(c)}(x_R)\Bigr)}{g_B\Big(\tilde{\Phi}_n(x_B)\Big)\,g^{(c)}_R\Bigl(\Phi^{(c)}_R(x_R)\Bigr)}\notag\\
&=\int \dd x_B\;b(x_B) 
+\sum_{c} \alpha_c
\int \dd x_B\,\dd x_R\;
r^{(c)}(x_B,x_R)\notag \\
&=\sum_{c} \alpha_c \int \dd x_B\,\dd x_R\;\left[b(x_B) 
+
r^{(c)}(x_B,x_R)\right] \; .
\label{eq:nlo_cs_mapped}
\end{align}
Each channel can be treated independently using \madnis and NCVs.
Denoting the variance of a single Monte Carlo evaluation in channel $c$ by $\text{Var}[\sigma^{(c)}_\text{NLO}]$, the statistical independence of the channels implies
\begin{align}
\text{Var}[\sigma_\text{NLO}]
=\sum_c \alpha_c^2\,
\text{Var}[\sigma^{(c)}_\text{NLO}]\;.
\label{eq:nlo_cs_channel_var_tot}
\end{align}
This gives the variance of the NLO integral from the individual channel variances.

\subsubsection*{NCVs for NLO}

The NLO cross section of Eq.\eqref{eq:nlo_cs_mapped} contains a single integrand per channel on the unit hypercube, with Born-like and real-emission contributions,
\begin{align}
f^{(c)}(x_B,x_R) = b(x_B) + r^{(c)}(x_B,x_R)\;.
\label{eq:nlo_combined}
\end{align}
This is exactly the nested structure of Eq.\eqref{eq:nested_split}, and both pieces can turn negative. 

The simplest option now treats each $f^{(c)}$ as the generic integrand of Section~\ref{sec:method_basic}. For each channel, we apply a single control variate $c^{(c)}_\theta(x)$ following Eq.\eqref{eq:cv_two_components} on $x=(x_B,x_R)$. We train it with the loss given in Eq.\eqref{eq:loss_single_ncv_total}. This removes the sign, but mixes the Born-like and real-emission contributions and does not exploit their conditional structure.

Alternatively, we keep the two integrals separate and follow the nested construction of Section~\ref{sec:method_nested}. For a given channel, we first introduce a conditional NCV over the radiation variables,
\begin{align}
\sigma^{(c)}_\text{NLO}
&=
\int \dd x_B\;
\Bigg[ \underbrace{
b(x_B) + A_\eta^{(c)}(x_B)
}_{\textstyle \equiv b^{(c)}_{\text{res},\eta}(x_B)} \Bigg]
+
\int \dd x_B\,\dd x_R\;
\Big[
\underbrace{r^{(c)}(x_B,x_R) - c_\eta^{(c)}(x_R| x_B) }_{\textstyle \equiv r^{(c)}_{\text{res},\eta}(x_B,x_R)}
\Big]\; ,
\label{eq:nlo_conditional_cv_phys}
\end{align}
with the NCV networks given as in Eqs.\eqref{eq:cv_conditional_def} and~\eqref{eq:cv_conditional_mass},
\begin{align}
c_\eta^{(c)}(x_R|x_B)
&= A_{+,\eta}^{(c)}(x_B)\,p_{+,\eta}^{(c)}(x_R| x_B)
- A_{-,\eta}^{(c)}(x_B)\,p_{-,\eta}^{(c)}(x_R| x_B)
\notag\\
\mwith \qquad
A_\eta^{(c)} &= A_{+,\eta}^{(c)}-A_{-,\eta}^{(c)}\; .
\label{eq:cv_conditional_phys_def}
\end{align}
Then, an unconditional NCV acts on the Born-like integral
\begin{align}
\sigma^{(c)}_\text{NLO}
&=
\int \dd x_B\;
\Big[
\underbrace{b^{(c)}_{\text{res},\eta}(x_B) - c^{(c)}_\theta(x_B)}_{\textstyle \equiv b^{(c)}_{\text{res},\theta,\eta}(x_B)}
\Big]
+ C^{(c)}_\theta
+
\int \dd x_B\,\dd x_R\;
r^{(c)}_{\text{res},\eta}(x_B,x_R) \; ,
\label{eq:nlo_two_cv_phys}
\end{align}
using a second, channel-dependent, unconditional NCV,
\begin{align}
c^{(c)}_\theta(x_B)
= C^{(c)}_{+,\theta}\,p^{(c)}_{+,\theta}(x_B)
- C^{(c)}_{-,\theta}\,p^{(c)}_{-,\theta}(x_B)
\qquad \mwith \qquad
C^{(c)}_\theta = C^{(c)}_{+,\theta}-C^{(c)}_{-,\theta}\; .
\label{eq:cv_unconditional_phys_def}
\end{align}
The conditional NCV is a learned function of the radiation variables at fixed Born-like kinematics. Its integral over the radiation variables $x_R$ is known and added back to the Born-like integrand. This is precisely the role of a subtraction term and its integrated counterpart in Eq.\eqref{eq:subtraction_def}, with the difference that it is not required to subtract a singularity. The original physics subtraction is reduced to its original purpose of regularizing the soft and collinear divergences, and the conditional NCV absorbs the remaining sign changes and the residual structure.
For sampling, both pieces are again combined to
\begin{align}
f^{(c)}_{\text{res},\theta,\eta}(x_B,x_R)
= b^{(c)}_{\text{res},\theta,\eta}(x_B)
+  r^{(c)}_{\text{res},\eta}(x_B,x_R)\;,
\label{eq:nlo_residual}
\end{align}
and sampled by a channel-dependent NIS flow $g^{(c)}_\phi$. The NCVs and the NIS are trained with the nested loss of Eq.\eqref{eq:loss_nested_total}, summed over channels,
\begin{align}
\loss_\text{nested}^\text{NLO} = \sum_c \loss_\text{nested}^{(c)}\;.
\label{eq:nlo_nested_loss_total}
\end{align}
The only NLO-specific modification concerns the training mixtures. The conditional NCV acts on the full $(x_B,x_R)$ space, the unconditional NCV on $x_B$ alone. We draw samples from the mixture density:
\begin{align}
q^{(c)}(x_B,x_R)
&= \frac14\Big[g^{(c)}_\text{neg} + g^{(c)}_\phi + \sum_{s=\pm} p^{(c)}_{s,\theta}(x_B)\,p_{s,\eta}^{(c)}(x_R|x_B)\Big]\Big|_\text{no-grad}
\notag\\
q^{(c)}(x_B)
&= \frac13\Big[ g^{(c)}_\text{neg} + \sum_{s=\pm} p^{(c)}_{s,\theta}(x_B) \Big]\Big|_\text{no-grad}\; .
\label{eq:nlo_mixtures}
\end{align}
The negativity barrier loss $\loss_\text{neg}$ only has gradients where a residual is negative. At NLO this is a small and narrow region that a density adapted to the full integrand does not populate. We therefore include $g^{(c)}_\text{neg}$, a \vegas grid adapted to the negative part $\max\big(0,-f^{(c)}\big)$ and re-adapted every $250$ steps to the negative part of the current residual. Together with the sampler $g^{(c)}_\phi$ and the NCV flows themselves, this gives a mixture that is broad enough for the shape and integral losses and at the same time resolves the regions where the negativity barrier is active.

\subsection{Top pair production with a gluon}
\label{sec:nlo_results1}

We first benchmark the NCV construction on the process 
\begin{align}
\Pep \Pem &\to \Pt \Ptbar \Pg \; .
\label{eq:born2}
\end{align}
The top mass regularizes the collinear divergences, leaving us with soft singularities, where the gluon can become unresolved and then requires a jet measurement function. We use the Durham jet algorithm~\cite{Catani:1991hj, Stagnitto:2025air} as a $k_T$-type clustering algorithm for $\Pep \Pem$ collisions to define the gluon jet. For every pair of particles $i$ and $k$ it evaluates
\begin{align}
    y_{ik} = \frac{\min\!\big(E_i^2, E_k^2\big)}{s}\, 2\,(1-\cos\theta_{ik})  \; .
\end{align}
This Durham measure becomes small exactly where the real emission becomes soft or collinear, so the jet boundary lies in the phase-space region affected by the dipole subtraction. For both processes we fix the collider energy and the jet resolution to
\begin{align}
    \sqrt{s} = 1~\text{TeV} \qquad \mand \qquad y_{\text{cut}} = 10^{-3} \; .
\end{align}
All Born, real-emission, and virtual amplitudes are generated with \mg~\cite{Alwall:2014hca,Frederix:2018nkq}.
We have verified that all setups reproduce the \mg cross section within the statistical uncertainties, so we can focus on their relative performance. We compare four setups
\begin{enumerate}
    \item \vegas
    \item \madnis
    \item single NCV (1-NCV $+$ \madnis)
    \item nested NCV (2-NCV $+$ \madnis)
\end{enumerate}
The hyperparameters can be found in Table~\ref{tab:nlo_hyperparameters}. For the NCVs, we normalize each positivity barrier by the negative component of the NCV acting on the corresponding residual, using $Z_{-,C}$ and $Z_{-,A}$ of Eq.\eqref{eq:nested_normalizations},
\begin{alignat}{9}
  Z_\text{neg} &= Z_\text{NIS} = Z^{(c)}_{C_-} = C^{(c)}_{-,\theta}\big|_\text{no-grad} 
  &&\qqqquad \text{(1-NCV)} \notag \\
  Z_{\text{neg},b} &= Z_\text{NIS} = Z^{(c)}_{C_-} 
  \qquad \text{and} \qquad Z_{\text{neg},r} = Z^{(c)}_{A_-} 
  &&\qqqquad \text{(2-NCV)} \; ,
  \label{eq:znorm_generic}
\end{alignat}
in direct analogy to the toy model in Eq.\eqref{eq:toy_norm_nested}.

Table~\ref{tab:ttg_results} shows that the \vegas and \madnis baseline come with a negative cross-section contribution of $4.2\%$, which originates from three distinct sources.
First, deep in the singular region $R$ and $S$ are both large and nearly equal, so their difference is dominated by floating-point noise of either sign, with weights far above the true size of $R_\text{fin}$. We remove them with a technical cut $|w|>10^3\,\langle w\rangle$, applied to all methods alike so the comparison stays fair.
Second, the measurement function can reject the real-emission configuration while accepting the dipole-mapped Born configuration. The subtraction term is then left uncancelled and enters with a negative sign. The Durham measure keeps this effect small, but some narrow negative spikes are unavoidable.
Third, away from the singular limits the dipoles are not required to approximate $R$, so their sum can exceed it. This genuine over-subtraction becomes the dominant source for the massless process of Eq.\eqref{eq:born_nlo}.

\begin{table}[t]
\centering
\begin{small}
\begin{tabular}{l|S[table-format=2.1]S[table-format=2.1]cS[table-format=1.3]S[table-format=2.1]}
\toprule
Method  & {$C_{+,\theta}/\sigma$ [\%]} & {$C_{-,\theta}/\sigma$ [\%]} & $\sigma_{\text{res}}^-/\sigma$ [\%] & {RV$_\sigma$} & {$\text{RV}_{\vegas}/\text{RV}$} \\
\midrule
\vegas  & {--}  & {--}  & 4.2 & 1.290 & {--} \\
\madnis & {--}  & {--} & 4.2 & 0.753 & 1.7  \\
1-NCV $+$ \madnis &  75.6 & 21.3 &  0.4 & 0.126 & 10.2  \\
2-NCV $+$ \madnis &  70.5 &  5.0 &  2.1 & 0.139 &  9.3  \\
\bottomrule
\end{tabular}
\end{small}
\caption{Performance metrics for $\Pep \Pem \to \Pt \Ptbar \Pg$ at NLO. All methods are evaluated for 6M phase-space points. We do not optimize the hyperparameters, so the numbers indicate the size of the effect rather than the optimum.
}
\label{tab:ttg_results}
\end{table}

\begin{figure}[t!]
    \includegraphics[width=0.49\linewidth]{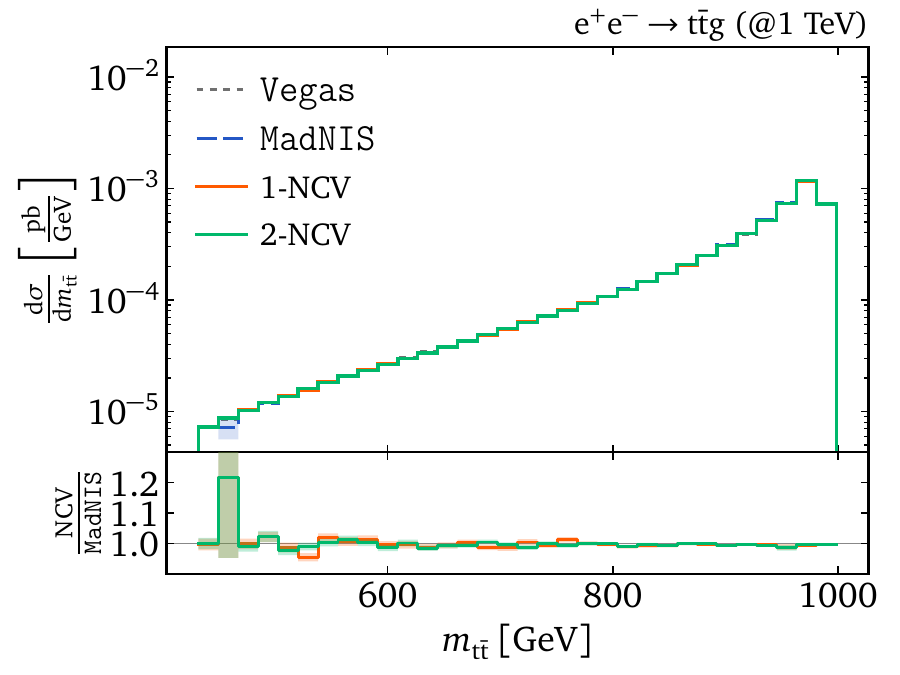}
    \includegraphics[width=0.49\linewidth]{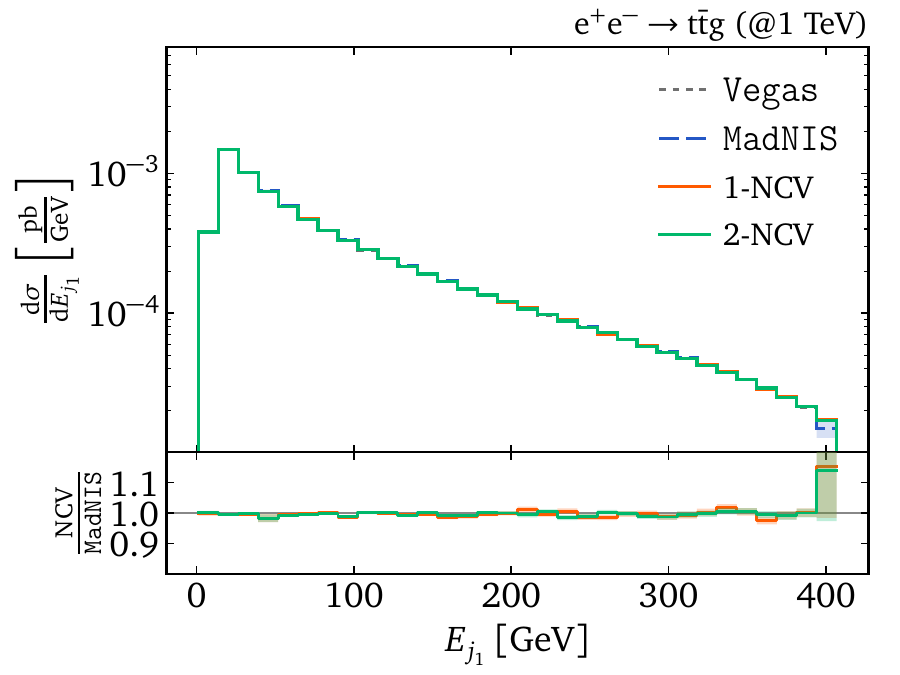}
    \caption{Differential distributions comparing all sampling methods for $\Pep\Pem\to\Pt \Ptbar \Pg$: invariant mass of the top pair $m_{\Pt\Ptbar}$ (left) and energy of the leading jet $E_{j_1}$ (right), evaluated for 6M phase-space points.}
    \label{fig:differential_histograms}
\end{figure}

We target all three sources of negative weights with the NCVs. For each of the six channels of Eq.\eqref{eq:channel_set_six} we train one NCV, giving us 6 NCVs and 6 NIS flows in the 1-NCV setup, and 12 NCVs and 6 NIS flows for the 2-NCV setup. We observe that \vegas handles this process reasonably well, because the integrand structure remains mild, while \madnis reduces the relative variance by roughly a factor two. 
Neither of them changes the integrand, so both keep the negative fraction $\sigma^-/\sigma=4.2\%$. Through Eq.\eqref{eq:rv_floor} this fraction implies a lower bound $\text{RV}_{\min}\simeq0.18$ which no pure importance sampling method can undercut. \madnis reaches $\text{RV}_\sigma=0.753$, a factor four above it, so the remaining variance is not yet limited by the sign but by the phase-space structure the sampler has to cover.

The 1-NCV construction removes the negative weights most effectively and reaches the lowest variance. 
The overshoot of $C_{-,\theta}$ is reabsorbed by $C_{+,\theta}$, leaving $\sigma^-_\text{res}/\sigma=0.4\%$, a negative-weight reduction of $90\%$, while reaching the lowest relative variance at $\text{RV}_\sigma = 0.126$, a factor $10.2$ ($6.0$) better than \vegas (\madnis). 
The 2-NCV setup performs equally well. Its $C_{-,\theta}$ is much smaller and, with $5\%$, close to the theoretical optimum of $4.2\%$, at the price of $\sigma^-_\text{res}/\sigma=2.1\%$ and a slightly larger variance.

Beyond the negative weights, the NCV also reduces the number of expensive integrand evaluations. Once trained, the net NCV contribution amounts to
\begin{align}
\frac{C_\theta}{\sigma} = \frac{C_{+,\theta}-C_{-,\theta}}{\sigma}
= 54\% \quad (\text{1-NCV})
\qquad \mand \qquad
65\% \quad (\text{2-NCV})
\end{align}
of the total cross section, sampled directly from the NCV flows with unit weights. Only the remaining $46\%$ and $35\%$ of the integral require an evaluation of the full integrand, including the expensive loop amplitudes. The NCV therefore combines the reduction of negative weights with a direct saving in computing time. On this metric the nested construction absorbs the larger fraction and needs roughly a quarter fewer evaluations of the full integrand than the single NCV.

Which of the two constructions is preferable depends on whether negative weights or matrix-element evaluations dominate the cost. In the toy models of Section~\ref{sec:method_toy} the nested construction was the better one throughout, which suggests that it might benefit from more optimization, \eg a different normalization or a different $\lambda_{\text{neg}}$.

In Figure~\ref{fig:differential_histograms} we show the invariant mass $m_{\Pt\Ptbar}$ and the energy of the first jet $E_{j_1}$ as  representative observables, where all methods agree within the statistical uncertainties. This confirms that we can use NCVs as a proper sampling method, so that differential distributions are reproduced bin by bin, just as for the LO application in Section~\ref{sec:lo_phys}.

\begin{figure}[t!]
    \includegraphics[width=0.495\linewidth]{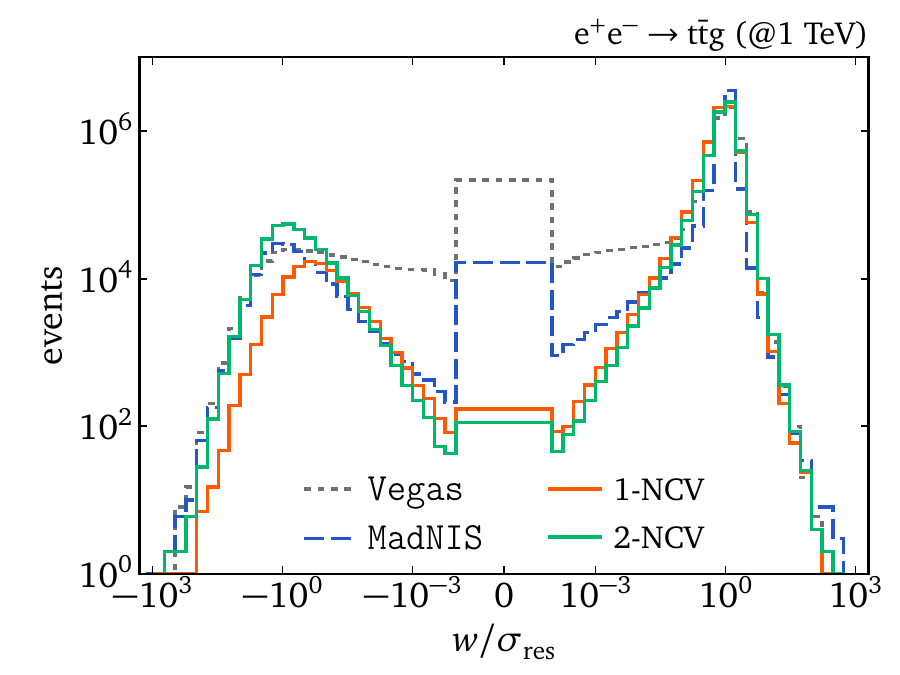}
    \includegraphics[width=0.495\linewidth]{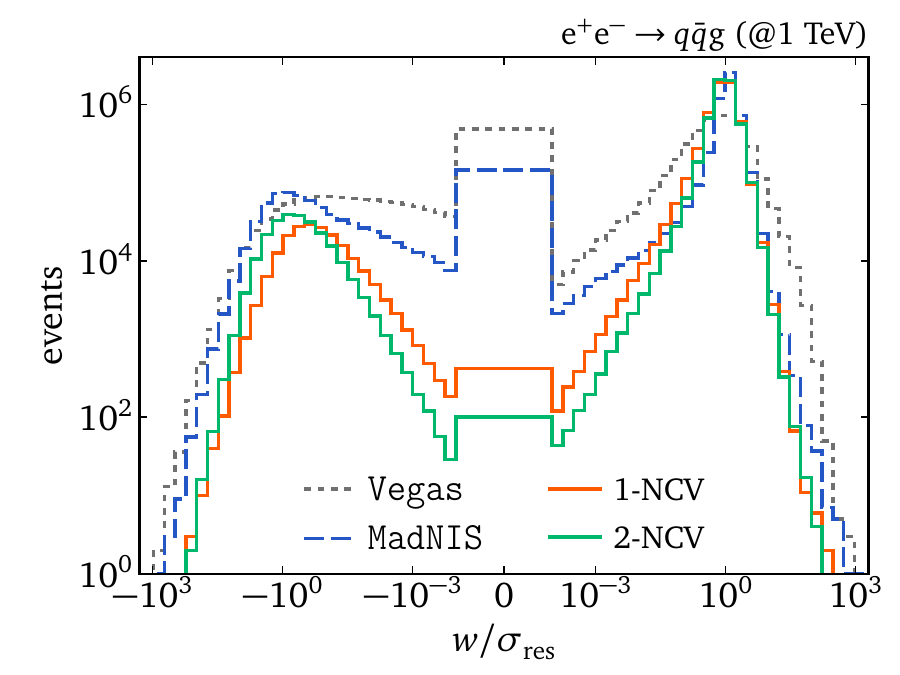}    
    \caption{Normalized weight distributions comparing all sampling methods for $\Pep\Pem\to\Pt \Ptbar \Pg $ (left) and $\Pep\Pem\to\Pq \Pqbar \Pg $ (right) evaluated for 6M phase-space points.}
    \label{fig:weight_distrubtion_NLO}
\end{figure}

The difference between the methods becomes apparent in the left panel of Figure~\ref{fig:weight_distrubtion_NLO}. For \vegas the weights spread over several orders of magnitude and the negative weights are almost uniformly distributed. \madnis narrows both peaks considerably, which is where its factor two in the relative variance comes from, but it leaves the negative contribution untouched. In an ideal scenario, we would observe two distinct peaks, a large one at positive weights and a very small one with some unavoidable negative weights. Only the NCVs approach this ideal: the positive peak sharpens further, so the residual weights concentrate around their mean and the variance is reduced, while the negative peak is not only lower but also shifted towards zero. The negative weights that remain are therefore smaller in magnitude, and their cancellation against the positive weights costs less effective statistics.

\begin{table}[b!]
\centering
\begin{small}
\begin{tabular}{l|S[table-format=2.1]S[table-format=2.1]S[table-format=2.1]S[table-format=1.3]c}
\toprule
Method  & {$C_{+,\theta}/\sigma$ [\%]} & {$C_{-,\theta}/\sigma$ [\%]} & {$\sigma_{\text{res}}^{-}/\sigma$ [\%]} & {RV$_\sigma$} & $\text{RV}_{\vegas}/\text{RV}$ \\
\midrule
\vegas  & {--} & {--} & 12.1 & 8.840 & {--}  \\
\madnis & {--} & {--} & 12.1 & 3.700 & 2.4  \\
1-NCV $+$ \madnis & 75.8 & 84.1 &  1.8 & 0.947 &  9.3 \\
2-NCV $+$ \madnis & 48.4 & 61.3 &  4.6 & 1.196 &  7.4  \\
\bottomrule
\end{tabular}
\end{small}
\caption{Performance metrics for $\Pep\Pem\to \Pq \Pqbar \Pg$ at NLO. All methods are evaluated for 6M phase-space points. We do not optimize the hyperparameters, so the numbers indicate the size of the effect rather than the optimum.
}
\label{tab:three_jet_results}
\end{table}

\subsection{Three-jet production}
\label{sec:nlo_results2}

Finally, we benchmark our NCVs on the massless process from Eq.\eqref{eq:born_nlo}
\begin{align}
\Pep \Pem &\to \Pq \Pqbar \Pg\; .
\label{eq:born3}
\end{align}
For $\Pt \Ptbar \Pg$ production the top mass regularizes the collinear limits, so only the gluon can become unresolved. Here all three final-state partons can become soft or collinear, so more dipoles contribute away from the singular limit, and the resulting over-subtraction nearly triples the negative-weight fraction.
This effect could be mitigated by using $\alpha$-restricted dipoles~\cite{Hasegawa:2014oya} that contribute only close to the divergent limits.

The comparison of the different methods follows the same setup as for $\Pt \Ptbar \Pg$ production, with six conditional NCVs the six integration channels. 
In Table~\ref{tab:three_jet_results}, we see that the 1-NCV reduces the negative contribution from $12.1\%$ to $1.8\%$, a reduction by $85\%$, and improves the relative variance by a factor $9.3$ ($3.9$) compared to \vegas (\madnis). The 2-NCV setup reduces the negative-weight contribution by $62\%$ reaching $\sigma^-_\text{res}/\sigma=4.6\%$, and reduces the relative variance by a factor of $7.4$ ($3.1$) compared to \vegas (\madnis).

In contrast to $\Pt \Ptbar \Pg$, the net NCV normalization is slightly negative, \ie $C_\theta/\sigma=-8\%$ and $-13\%$, because the deep negative structures require a large $C_{-,\theta}$. The NCV therefore acts mainly as a lift of the negative region, and the largest gain is a reduction of negative weights and of the variance rather than a saving in matrix-element evaluations.

As before, we have checked that the kinematic distributions agree between all methods within the statistical uncertainties.
The advantage of the NCVs is again visible in the right panel of Figure~\ref{fig:weight_distrubtion_NLO}, with the same range reduction and removal of negative weights as for $\Pt \Ptbar \Pg$ production.

\section{Outlook}

Neural control variates provide a simple and general handle on two leading limitations of Monte Carlo event generation, a wide range of weights and negative weights. A learned control variate reshapes the integrand, flattens its bulk for range reduction, and lifts its negative part for sign removal. Additional neural importance sampling further reduces the variance of the residual, while the control-variate integral produces trivially unweighted events. Crucially, our construction is designed for event generation, not just integration~\cite{Shyamsundar:2023jtz}.

We first applied this combined NCV+NIS setup to the process $\Pg\Pg\to\Pt\Ptbar\Pg\Pg$ at LO. The range reduction improved the unweighting efficiency by a factor of $2.7$ relative to \madnis alone. This improvement should increase for more complex processes. 

At NLO, we compared a single joint NCV with the nested construction, combining a conditional NCV in the radiation variables with an unconditional NCV in the Born-like variables.
After removing the real-emission divergences, for instance through the usual Catani--Seymour dipoles, the conditional NCV can be viewed as an optimized finite soft and collinear subtraction term. Our setup is independent of the subtraction scheme and only requires a regularization.

We benchmarked our setup for two processes, $\Pep \Pem \to \Pt \Ptbar \Pg$ and $\Pep \Pem \to \Pq \Pqbar \Pg$. In both cases, the NCVs improved the variance and reduced the negative contributions. They reduced the relative variance by roughly an order of magnitude compared to \vegas and the negative contribution to the total cross section from $4.2\%$ to $0.4\%$ and from $12.1\%$ to $1.8\%$. For $\Pt \Ptbar \Pg$ production the NCV also absorbs $54\%$ to $65\%$ of the cross section into the trivially sampled NCV contribution, so only a third to a half of the integral requires an evaluation of the expensive loop amplitudes. These numbers are not the results of an intensive hyperoptimization and are only meant to illustrate the power of NCVs for realistic physics problems.

\subsection*{Acknowledgments}

We thank Raoul Röntsch for valuable discussions on subtraction schemes and negative weights. We are very grateful to Michael Kr\"amer, Kirill Melnikov, and Thomas Gehrmann for their encouraging feedback. We thank Giovanni De Crescenzo for useful discussions about the NLO implementation in MadGraph and Daniel Schiller for help with the MadAgents. We acknowledge support by the Deutsche Forschungsgemeinschaft (DFG, German Research Foundation) under grant 396021762 -- TRR~257: \textsl{Particle Physics Phenomenology after the Higgs Discovery}. TH is supported by the PDR-Weave grant FNRS-DFG numéro T019324F (40020485), and by FRS-FNRS (Belgian National Scientific Research Fund) IISN projects 4.4503.16 (MaxLHC). SV is funded by the Carl-Zeiss-Stiftung through the project \textit{Model-Based AI: Physical Models and Deep Learning for Imaging and Cancer Treatment}.

\clearpage
\appendix
\section{Architectures and hyperparameters}
\label{app:arch}

\begin{table}[h!]
\centering
\begin{small}
\begin{tabular}{llc}
\toprule
Component & hyperparameter & Value \\
\midrule
\multirow{6}{*}{all}
 & optimizer & Adam \\
 & learning rate & $2\times10^{-3}$ \\
 & batch size & $4096$ \\
 & scheduler & cosine annealing \\
 & training iterations & $3000$ \\
 & evaluation sample & $10^6$ points \\
\midrule
\multirow{5}{*}{NCV}
 & $\lambda_\text{int}$ & $1$ \\
 & $\lambda_\text{neg}$ & $1$ \\
 & $\beta_\text{max}$ & $10$, annealed over 900 iterations \\
 & normalizing flows & 3 layers, 16 bins, 64 hidden dim \\
 & normalizations $A_{\pm,\eta}(x_B)$ & 3 layers, 48 hidden dim, softplus \\
\midrule
\madnis
 & normalizing flow & 3 layers, 32 bins, 64 hidden dim \\
\bottomrule
\end{tabular}
\end{small}
\caption{
Hyperparameters for the toy study of Section~\ref{sec:method_toy}.}
\label{tab:toy_hyperparameters}
\end{table}

\begin{table}[h!]
\centering
\begin{small} 
\begin{tabular}{llc}
\toprule
Component & hyperparameter & Value \\
\midrule
\multirow{6}{*}{all}
 & optimizer & Adam \\
 & learning rate & $10^{-3}$ \\
 & batch size & $1024$ \\
 & scheduler & cosine annealing \\
 & training iterations & $10000$ \\
 & evaluation sample & $10^5$ points \\
 \midrule
 \multirow{5}{*}{NCV}
& $\lambda_\text{int}$ & $10$ \\
& $\lambda_\text{cut}$ & $1$ \\
& $\lambda_\text{neg}$ & $0$ or $10$ \\
& normalizing flow & 4 layers, 10 bins, 256 hidden dim \\
& $\beta(t)=\text{const}$ & $3$ \\
\midrule
\madnis & normalizing flow & 3 layers, 10 bins, 128 hidden dim \\
\bottomrule
\end{tabular}
\end{small} 
\caption{
Hyperparameters for the LO study of Section~\ref{sec:lo_phys}.}
\label{tab:lo_hyperparameters}
\end{table}

\begin{table}[t]
\centering
\begin{small}
\begin{tabular}{llc}
\toprule
Component & hyperparameter & Value \\
\midrule
\multirow{7}{*}{all}
 & optimizer & Adam \\
 & learning rate (flows / normalizations) & $10^{-3}$ / $5\times10^{-3}$ \\
 & batch size & $8192$ \\
 & scheduler & cosine annealing \\
 & training iterations & 36k \\
 & weight cut & $|w| > 10^3\,\langle w\rangle$ \\
 & evaluation sample & $6\times10^6$ points \\
\midrule
\multirow{5}{*}{NCV}
 & $\lambda_\text{int}$ & $1$ \\
 & $\lambda_\text{neg}$ &$0.5$  \\
 & $\beta_\text{max}$ & $30$, annealed over 10k iterations \\
 & normalizing flows  & 3 layers, 32 bins, 128 hidden dim \\
 & normalizations $A_{\pm,\eta}(x_B)$ & 3 layers, 128 hidden dim, softplus \\
\midrule
\multirow{3}{*}{\madnis}
 & normalizing flow & 3 layers, 32 bins, 128 hidden dim \\
 & fine-tune iterations & $6000$, frozen NCV \\
 & fine-tune learning rate & $3\times10^{-4}$ \\
\bottomrule
\end{tabular}
\end{small}
\caption{Hyperparameters for the NLO study of Section~\ref{sec:nlo_results1}.}
\label{tab:nlo_hyperparameters}
\end{table}

\clearpage
\bibliography{tilman,refs}

\end{document}